\documentclass[manuscript]{aastex}
\usepackage{bm}

\shorttitle{CR Decrease with IP-shock Observed with GMDN}
\shortauthors{Kozai et al.}

\begin{document}


\title{Average spatial distribution of cosmic rays\\
   behind the interplanetary shock\\
-- Global Muon Detector Network observations --}


\author{M. Kozai\altaffilmark{1}, K. Munakata\altaffilmark{1}, C. Kato\altaffilmark{1}, T. Kuwabara\altaffilmark{2}, M. Rockenbach\altaffilmark{3},
A. Dal Lago\altaffilmark{4}, N. J. Schuch\altaffilmark{3}, C. R. Braga\altaffilmark{4}, R. R. S. Mendon\c{c}a\altaffilmark{4},
H. K. Al Jassar\altaffilmark{5}, M. M. Sharma\altaffilmark{5}, M. L. Duldig\altaffilmark{6},
J. E. Humble\altaffilmark{6}, P. Evenson\altaffilmark{7}, I. Sabbah\altaffilmark{8}, and M. Tokumaru\altaffilmark{9}}
\email{13st303f@shinshu-u.ac.jp} \and \email{kmuna00@shinshu-u.ac.jp}
\altaffiltext{1}{Department of Physics, Shinshu University, Matsumoto, Nagano 390-8621, Japan.}
\altaffiltext{2}{Graduate School of Science, Chiba University, Chiba City, Chiba 263-8522, Japan.}
\altaffiltext{3}{Southern Regional Space Research Center (CRS/INPE), P.O. Box 5021, 97110-970, Santa Maria, RS, Brazil.}
\altaffiltext{4}{National Institute for Space Research (INPE), 12227-010 S\~{a}o Jos\'{e} dos Campos, SP, Brazil.}
\altaffiltext{5}{Physics Department, Kuwait University, P.O. Box 5969 Safat, Kuwait 13060.}
\altaffiltext{6}{School of Physical Sciences, University of Tasmania, Hobart, Tasmania 7001, Australia.}
\altaffiltext{7}{Bartol Research Institute and Department of Physics and Astronomy, University of Delaware, Newark, DE 19716, USA.}
\altaffiltext{8}{Department of Natural Sciences, College of Health Sciences, Public Authority of Applied Education and Training, Kuwait City 72853, Kuwait.}
\altaffiltext{9}{Institute for Space-Earth Environmental Research, Nagoya University, Nagoya, Aichi 464-8601, Japan.}



\begin{abstract}
   We analyze the galactic cosmic ray (GCR) density and its spatial gradient in Forbush Decreases (FDs) observed
   with the Global Muon Detector Network (GMDN) and neutron monitors (NMs).
   By superposing the GCR density and density gradient observed in FDs
   following 45 interplanetary shocks (IP-shocks),
   each associated with an identified eruption on the sun,
   we infer the average spatial distribution of GCRs behind IP-shocks.
   We find two distinct modulations of GCR density in FDs,
   one in the magnetic sheath and the other in the coronal mass ejection (CME) behind the sheath.
   The density modulation in the sheath is dominant in the western flank of the shock,
   while the modulation in the CME ejecta stands out in the eastern flank.
   This east-west asymmetry is more prominent in GMDN data responding
   to $\sim 60$ GV GCRs than in NM data responding to $\sim 10$ GV GCRs,
   because of softer rigidity spectrum of the modulation in the CME ejecta than in the sheath.
   The GSE-$y$ component of the density gradient, $G_y$ shows a negative (positive) enhancement in FDs caused by the eastern (western) eruptions,
   while $G_z$ shows a negative (positive) enhancement in FDs by the northern (southern) eruptions.
   This implies the GCR density minimum being located behind the central flank of IP-shock
   and propagating radially outward from location of the solar eruption.
   We also confirmed the average $G_z$ changing its sign above and below the heliospheric current sheet,
   in accord with the prediction of the drift model for the large-scale GCR transport in the heliosphere.
\end{abstract}


\keywords{astroparticle physics, cosmic rays, interplanetary medium, methods: data analysis, solar wind, Sun: coronal mass ejections (CMEs)}



\section{Introduction}\label{sec:Int}
Short term decreases in the galactic cosmic ray (GCR) isotropic intensity (or density)
following geomagnetic storm sudden commencements (SSCs) were first observed by \citet{For37} (Forbush Decreases, FDs).
In general, FDs start with a sudden decrease within 3 hours of the SSC onset \citep{Loc60},
reach maximum depression within about a day and recovers to the usual level over several days (recovery phase).
Most of the decreases follow geomagnetic SSCs but correlation studies between
the ground-based cosmic ray data and spacecraft \citep[e.g.][]{Fan60} or solar radio \citep[e.g.][]{Oba62} data
indicate that the origin of the FD is not the geomagnetic storm but the interplanetary shock (IP-shock)
associated with the solar eruption such as the coronal mass ejection (CME),
which causes the SSC as well \citep{Yer06,Gop07}.\par
The depleting effect of IP-shocks on GCRs is explained by the ``propagating diffusive barrier'' model \citep{Wib98}.
The compressed and disturbed magnetized plasma in the sheath behind the IP-shock reduces the GCR diffusion from the outer heliosphere
due to the enhanced pitch angle scattering and works as a diffusive barrier.
The diffusive barrier suppresses the inward flow arising from the radial density gradient of GCRs
and sweeps out GCRs as it propagates radially outward, forming the GCR depleted region behind the IP-shock.\par
Investigating a relation between the heliographic longitude of associated solar eruptions on the sun and the magnitude of GCR depression in FDs,
a number of studies suggest the east-west asymmetry (E-W asymmetry) of FDs
associated with eruptions on the eastern region of the sun have slightly larger magnitude
than western eruptions \citep{Kam61,Sin62,Yos65,Hau65,Bar73a,Bar73b,Can96}.
It is also reported that large FDs with prominent magnitudes are often observed in association with
eruptions near the central meridian of the sun. \citet{Yos65} called this the ``center-limb effect''.
We note, however, that the E-W asymmetry presented by previous papers seems insignificant
due to a large event-by-event dispersion of the maximum density depression in FD masking the systematic E-W dependence.\par
\citet{Bar73a,Bar73b} and \citet{Can00} gave a comprehensive interpretation of the observations including the E-W asymmetry and center-limb effect
applying the magnetic configuration model of \citet{Hun72} to FDs.
The IP-shocks associated with solar eruptions are driven by the ejected ``driver gas'' \citep{Hir70}, i.e. the interplanetary CME.
The central region of the CME (or the CME ejecta), whose longitudinal extent is less than 50$^\circ$ at 1 AU \citep{Can03},
is detected only for IP-shocks originating near the central meridian,
while the accompanying shock formed ahead of the CME has a greater longitudinal extent exceeding 100$^\circ$ \citep{Can88}.
A closed magnetic field configuration called the magnetic flux rope (MFR)
is formed in the central region of the CME \citep{Bur81,Kle82}.
Expansion of the MFR excludes GCRs from penetrating into the MFR, causing a prominent FD as found by \citet{Can96}.
The E-W asymmetry, on the other hand, is attributed to the IP-shock which has a global effect on the GCRs \citep{Can94}.
The interplanetary magnetic field (IMF) has a spiral configuration known as the Parker spiral \citep{Par58} and
the eruption site on the solar photosphere moves toward west due to the sun's rotation before the IP-shock arrives at Earth.
The compressed IMF in the sheath of IP-shock, therefore, has a larger magnitude at the western flank of the IP-shock than at the eastern flank,
leading to a small diffusion coefficient of the GCR pitch angle scattering \citep{Jok71} and a larger FD in the eastern events.
This CME-driven shock model is also consistent with the observed longitudinal distribution of the solar energetic particles \citep{Rea95,Rea96}.\par
In addition to the temporal variation of GCR density, FDs are often accompanied by dynamic variations of the anisotropic intensity of GCRs (or GCR anisotropy)
observed with ground-based detectors such as neutron monitors and muon detectors.
The cosmic ray counting rate observed with a ground-based detector is known to show a diurnal variation \citep{Hes36},
indicating an equatorial GCR flow from the direction of the local time when a maximum count rate is observed.
The enhancement of amplitude and the rotation of phase of the diurnal variation accompanying FDs were first reported by \citet{Dug62} and
\citet{Wad80} performed a statistical analysis of the evolution of diurnal anisotropy for SSC events.
\citet{Dug70} and \citet{Sud81} also found enhanced north-south asymmetry in GCR intensities
observed with the northern and southern geographic polar detectors, indicating an enhancement of the north-south GCR anisotropy in FDs.
Combination of the observed diurnal and north-south anisotropies enabled \citet{Nag68} to infer the three-dimensional density distribution.
However, after that, such a three-dimensional analysis of the transient anisotropy was rarely performed until a worldwide detector network started operation.
The counting rate of a single neutron monitor, which is analyzed in most previous studies,
contains contributions from the GCR density and anisotropy superposed to each other and analyzing these two contributions separately has been difficult.
Also the analysis of the diurnal variation provides only the daily mean of the equatorial anisotropy,
which is insufficient for analyzing the dynamic variation during FDs.
This has been a problem also in analysis of the temporal variation of GCR density in previous studies, as pointed out by \citet{Can96}.\par
In this paper, we put a special emphasis on the analysis of the anisotropy because most of the former works on FDs analyzed only the GCR density.
The first order anisotropy corrected for the solar wind convection represents a GCR flow proportional to the spatial density gradient of GCRs.
We can thus derive the density gradient from the observed anisotropy based on Parker's transport equation of GCRs in the heliosphere \citep{Par65}.
While the scalar density only reflects the local information at the observation point,
the density gradient vector allows us to infer the three-dimensional spatial distribution of GCRs behind the IP-shock.
Only a worldwide detector network viewing various directions in space simultaneously can observe the GCR density and anisotropy separately each with a sufficient temporal resolution.
The Global Muon Detector Network (GMDN), which is capable of measuring the isotropic intensity and three-dimensional anisotropy of $\sim 60$ GV GCRs on an hourly basis,
was completed with four multi-directional muon detectors at Nagoya (Japan), Hobart (Australia), S\~ao Martinho da Serra (Brazil), and Kuwait University (Kuwait) in 2006.
An analysis method of deducing the GCR density and anisotropy from the GMDN data has been developed \citep{Oka08,Fus10a}.\par
In former analyses of the IP-shock events observed with the GMDN,
the GCR density and density gradient have been used to
analyze a geometry of the GCR depleted region in each individual FD \citep{Mun03,Mun06,Kuw04,Kuw09,Roc14}.
In this paper, we perform superposed epoch analyses of the GCR density and gradient derived from observations with the GMDN for 45 IP-shock events
and analyze for the first time the average spatial distribution of GCR density behind the IP-shock.\par
The derivation of the GCR anisotropy, density, and the density gradient from the GMDN data is explained in Sections \ref{sec:AnaXi} and \ref{sec:AnaG}.
We describe our method of identifying the IP-shock arrivals and the associated CMEs in Section \ref{sec:IdEv}.
After viewing three event samples in Section \ref{sec:Samp},
we perform superposition analyses of the density and gradient in Section \ref{sec:ss} and
deduce the average spatial distribution of GCR density behind IP-shock.
In Section \ref{sec:Sum}, we present the summary and conclusions.

\section{Data analysis}
\subsection{Derivation of the first order anisotropy and the density}\label{sec:AnaXi}
We analyze a percent deviation from the 27 days average of the pressure corrected hourly count rate,
$I_{i,j}(t)$ of muons in the $j$-th directional channel of the $i$-th detector in the GMDN at the universal time $t$.
Detail descriptions of the GMDN and data analyses can be found in \citet{Oka08}.
Three components $\left( \xi^{\rm GEO}_{x}(t), \xi^{\rm GEO}_{y}(t), \xi^{\rm GEO}_{z}(t) \right)$ of the first order anisotropy in the geographic (GEO) coordinate system
are derived by best-fitting the following model function to $I_{i,j}(t)$.
\begin{eqnarray}
   I^{fit}_{i,j}(t) = I_0(t)c_{0 i,j}^0 &+& \xi^{\rm GEO}_x(t)(c_{1 i,j}^1 \cos \omega t_i - s_{1 i,j}^1 \sin \omega t_i) \nonumber \\
   &+& \xi^{\rm GEO}_y(t)(s_{1 i,j}^1 \cos \omega t_i + c_{1 i,j}^1 \sin \omega t_i) \nonumber \\
   &+& \xi^{\rm GEO}_z(t) c_{1 i,j}^0
   \label{eq:fit}
\end{eqnarray}
where $I_0(t)$ is a parameter representing contributions from the GCR density and the atmospheric temperature effect,
$t_{i}$ is the local time in hours at the $i$-th detector, $c^{0}_{0 i,j}$, $c^{1}_{1 i,j}$, $s^{1}_{1 i,j}$ and $c^{0}_{1 i,j}$ are the coupling coefficients and $\omega=\pi/12$.
The coupling coefficients are calculated using the response function of atmospheric muon intensity to primary cosmic rays \citep{Nag71,Mur79,Fuj84}.
In this calculation, we assume a rigidity independent anisotropy with the upper limit rigidity set at $10^{5}$ GV, far above the most responsive rigidity of the muon detectors.
We additionally apply an analytical method developed for removing
the atmospheric temperature effect from the derived anisotropy \citep[see Appendix A1 of][]{Oka08}.
The derived anisotropy vector in the GEO coordinate system is then transformed
to the geocentric solar ecliptic (GSE) coordinate system.\par
The analytical method of \citet{Oka08} removes the temperature effect from the derived anisotropy,
but not from the density.
We derive the GCR density $I_0(t)$ separately from the anisotropy derivation by best-fitting the model function (\ref{eq:fit}) to $I_{i,j}(t)$ in this paper,
on a simple assumption that the temperature effect should be almost averaged out in this best-fitting to all GMDN stations at various locations around the world.
In order to evaluate how the derived $I_0(t)$ is influenced by the temperature effect,
we performed the following analyses.
By using the Global Forecast System (GFS) model\footnote{http://www.emc.ncep.noaa.gov}
for the vertical distribution of high altitude atmospheric temperature,
one year GMDN data ($I_{i,j}(t)$) in 2009 was corrected for the temperature effect on an hourly basis
(\citet{Ber12} and Dr. V. Yanke, private communication).
One year data in 2009 during the last solar activity minimum was chosen for our analysis
to minimize possible technical influences to the correction from FDs in the GMDN data,
while the performance of an ideal correction method should be independent of the solar activity.
We then obtained a Gaussian-like distribution of the difference ($\Delta I_0(t)$)
between $I_0(t)$s derived from $I_{i,j}(t)$s before and after the correction.
We confirmed that the yearly mean $\Delta I_0(t)$ is 0.00 \%,
while a few \% seasonal variation due to the temperature effect is recorded in the uncorrected $I_{i,j}(t)$.
We also found that the standard deviation of $\Delta I_0(t)$ is 0.18 \%.
We will use this value as a measure of the temperature effect included in $I_0(t)$ in Section \ref{sec:ssDen}.
We derive $I_0(t)$ from $I_{i,j}(t)$ uncorrected for the temperature effect in this paper,
while a fully reliable correction process of $I_{i,j}(t)$
which will allow us to derive $I_0(t)$ free from the temperature effect
is under development.\par
The GCR density variation free from the temperature effect can be deduced from count rates recorded by polar neutron monitors (NMs), as
\begin{equation}
   I_0^{\rm NM}(t) = \frac{I^{\rm Thule}(t) + I^{\rm McMurdo}(t)}{2}
   \label{eq:NM}
\end{equation}
where $I^{\rm Thule}(t)$ and $I^{\rm McMurdo}(t)$ are percent deviations from the 27 days averages of
the pressure corrected hourly count rates recorded by the Thule and McMurdo NMs in Greenland and Antarctica, respectively.
The $I_0^{\rm NM}(t)$ in equation (\ref{eq:NM}) gives a good measure of the GCR density,
also because it contains only minor effects of the diurnal and north-south anisotropies \citep{Sud81}.
By comparing $I_0(t)$ by the GMDN with $I_0^{\rm NM}(t)$ by NMs in Section \ref{sec:Samp},
we will confirm that our conclusions in this paper are not seriously affected
by the atmospheric temperature effect.
Since the median rigidity of primary GCRs observed by NMs is $\sim 10$ GV,
while the median rigidity of GCRs observed by the GMDN is $\sim 60$ GV,
we can also analyze the rigidity dependence of the GCR density depression in FDs by comparing $I_0(t)$ with $I_0^{\rm NM}(t)$.

\subsection{Derivation of the density gradient vector}\label{sec:AnaG}
The first order anisotropy vector derived from equation (\ref{eq:fit}) is expressed in terms of the spatial density gradient, as \citep{Gle69}
\begin{equation}
   {\bm \xi}(t) = - \frac{3\bm{S}}{vU} = \frac{3}{vU} {\bf K} \cdot \nabla U - \frac{2 + \gamma}{v}(\bm{V}_{\rm SW} - \bm{V}_{E})
   \label{eq:flow}
\end{equation}
where $U$ is the GCR density, $\bm{S}$ is the bulk flow vector of GCRs,
${\bf K}$ is the diffusion tensor representing the diffusion and drift effects of GCRs,
$\gamma$ is the power-law index of the GCR energy spectrum \citep{Com35,Gle68},
$\bm{V}_E$ is the velocity of Earth's orbital motion around the sun,
${\bm V}_{\rm SW}$ is the solar wind velocity vector,
and $v$ is the particle speed,
which is approximately equal to the speed of light for GCRs observed by the GMDN and NMs.
The index $\gamma$ is set at 2.7 referring to \citet{Mur79}
who calculated the muon response function used for calculating the coupling coefficients in this paper.
The anisotropy vector ${\bm \xi}$ in equation (\ref{eq:flow}) is defined to direct opposite to ${\bm S}$, pointing toward the upstream direction of ${\bm S}$.
We correct the anisotropy vector for the solar wind convection and the Compton-Getting effect,
using the solar wind velocity ${\bm V}_{\rm SW}$ in spacecraft data and Earth's orbital motion speed $V_{\rm E}$ set at 30 km/s.
Hourly solar wind velocity ${\bm V}_{\rm SW}(t)$ for our analysis is mainly given by the ACE level 2 data\footnote{http://www.srl.caltech.edu/ACE/ASC/}
and we also use the WIND spacecraft data\footnote{http://wind.nasa.gov/data.php} when there is a gap in the ACE data, after confirming consistency between two data sets before and after the data gap.
The ACE and WIND data are lagged for 1 hour as a rough correction for the solar wind transit time between the spacecraft at the L1 Lagrangian point and Earth.
The corrected anisotropy ${\bm \xi}^w(t)$ is related to the spatial gradient of the GCR density at Earth, ${\bm G}(t) = \nabla U/U$ as
\begin{equation}
   \bm{\xi}^w(t) = \frac{3}{v} {\bf K} \cdot \bm{G} =
   R_L(t) \left\{ \alpha_\parallel \bm{G}_\parallel(t) + \alpha_\perp \bm{G}_\perp(t) -\frac{\bm{B}(t)}{B(t)} \times \bm{G}_\perp(t) \right\} 
   \label{eq:xg}
\end{equation}
where ${\bm G}_\parallel(t)$ and ${\bm G}_\perp(t)$ are the density gradient components parallel and perpendicular to the IMF,
${\bm B}(t)$ is the IMF vector in the ACE or WIND data lagged for 1 hour, and $R_L(t)$ is the Larmor radius of GCR particles.
The $\alpha_\parallel$ and $\alpha_\perp$ are dimensionless mean free paths of the GCR pitch angle scattering, defined as
\begin{equation}
   \alpha_\parallel = \lambda_\parallel(t) / R_L(t) {\rm \ and\ } \alpha_\perp = \lambda_\perp(t) / R_L(t)
\end{equation}
where $\lambda_\parallel$ and $\lambda_\perp$ are the parallel and perpendicular mean free paths.
From equation (\ref{eq:xg}), the density gradient ${\bm G}(t)$ is given in terms of the anisotropy, as
\begin{equation}
   {\bm G}(t) = {\bm G}_\parallel(t) + {\bm G}_{\perp}(t) =
   \frac{1}{R_L \alpha_\parallel} {\bm \xi}_\parallel^w + \frac{1}{R_L (1 + \alpha_\perp^2)} \left( \alpha_\perp {\bm \xi}_\perp^w + \frac{\bm B}{B} \times {\bm \xi}_\perp^w \right)
   \label{eq:g}
\end{equation}
where ${\bm \xi}_\parallel^w(t)$ and ${\bm \xi}_\perp^w(t)$ are the anisotropy components parallel and perpendicular to the IMF.
The $R_L(t)$ is calculated as $R_L(t) = P/cB(t)$ with $c$ denoting the speed of light and
$P$ denoting the rigidity of GCR particles that we set at 60 GV, the representative median rigidity of primary GCRs observed with the GMDN.
Following theoretical calculations by \cite{Bie04}, we assume in this paper constant $\alpha_\parallel$ and $\alpha_\perp$
at $\alpha_\parallel = 7.2$ and $\alpha_\perp = 0.05 \alpha_\parallel$.
This assumption is also used by \citet{Oka08} and \citet{Fus10a} and proved
to result in a reasonable GCR density distribution in the vicinity of the interplanetary disturbance.
Moreover, \citet{Fus10b} deduced the parallel mean free path $\lambda_\parallel$ from the observed ``decay length'' of the loss-cone precursor of an IP-shock event
and obtained $\lambda_\parallel$ comparable to our assumption of $\lambda_\parallel = 7.2R_L$.

\subsection{Identification of IP-shocks associated with solar eruptions}\label{sec:IdEv}
We infer the spatial distribution of GCRs behind IP-shocks
by analyzing temporal variations of the GCR density and its spatial gradient in IP-shock events, each identified with a source location on the sun.
IP-shocks are known to cause the geomagnetic SSCs in general \citep{Smi83,Wan06}.
We identify IP-shock arrivals with SSCs listed by the German Research Centre for Geosciences (GFZ) and extract 79 CME-associated shocks (CME events) from 214 SSCs in a period between 2006 and 2014,
referring to the space weather news (SW news) of the National Institute of Technology (NIT), Kagoshima College\footnote{http://www.kagoshima-ct.ac.jp/} on the date of each SSC occurrence.
The SW news reports the current status of the solar surface and interplanetary space each day, monitoring SDO, SOHO, ACE, and GOES spacecraft data,
geomagnetic indices and solar wind prediction by the Space Weather Prediction Center (SWPC), NOAA.
It estimates not only the interplanetary origin of each geomagnetic storm but also the associated solar event,
allowing us to associate a CME eruption on the sun with each IP-shock event recorded at Earth.
For the heliographic location of the CME eruption on the solar surface,
we use the location of the associated H-$\alpha$ flare or filament disappearance in the solar event listed by SWPC.\par
\begin{figure}
   \epsscale{0.5}
   \plotone{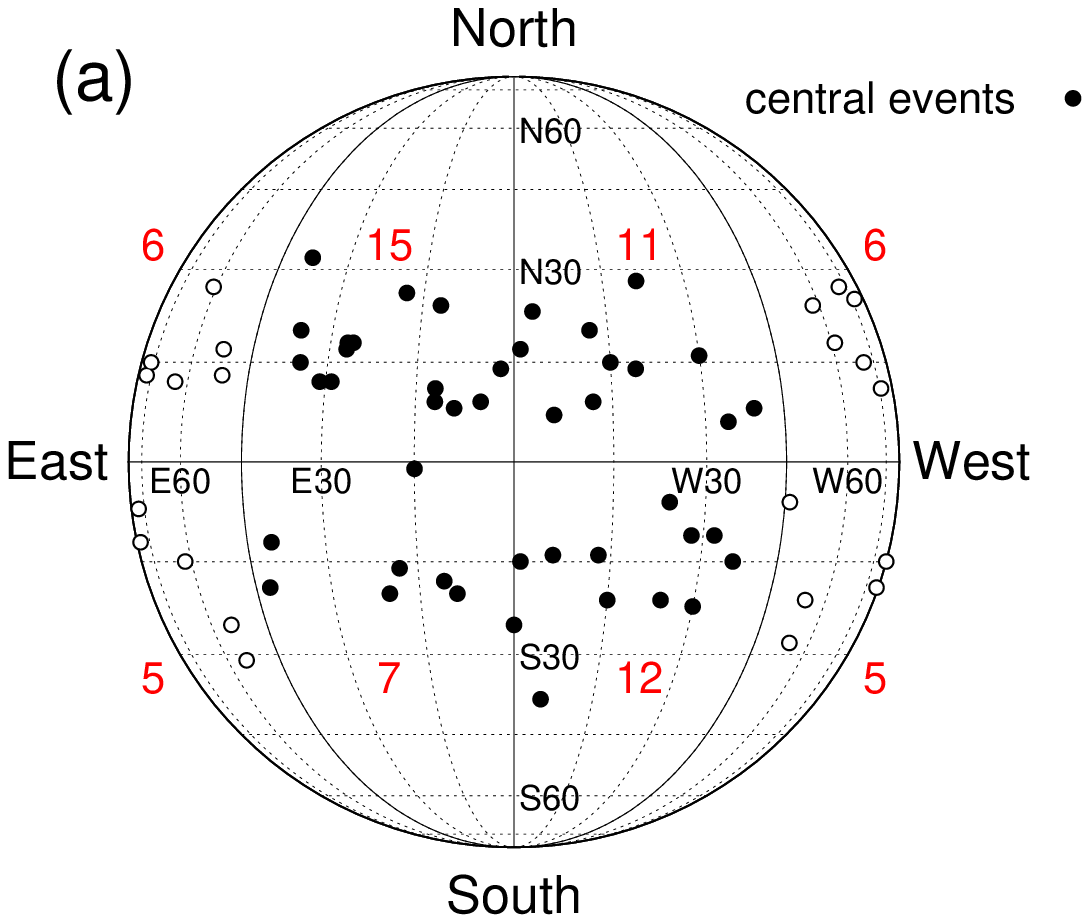}
   \epsscale{0.8}
   \plotone{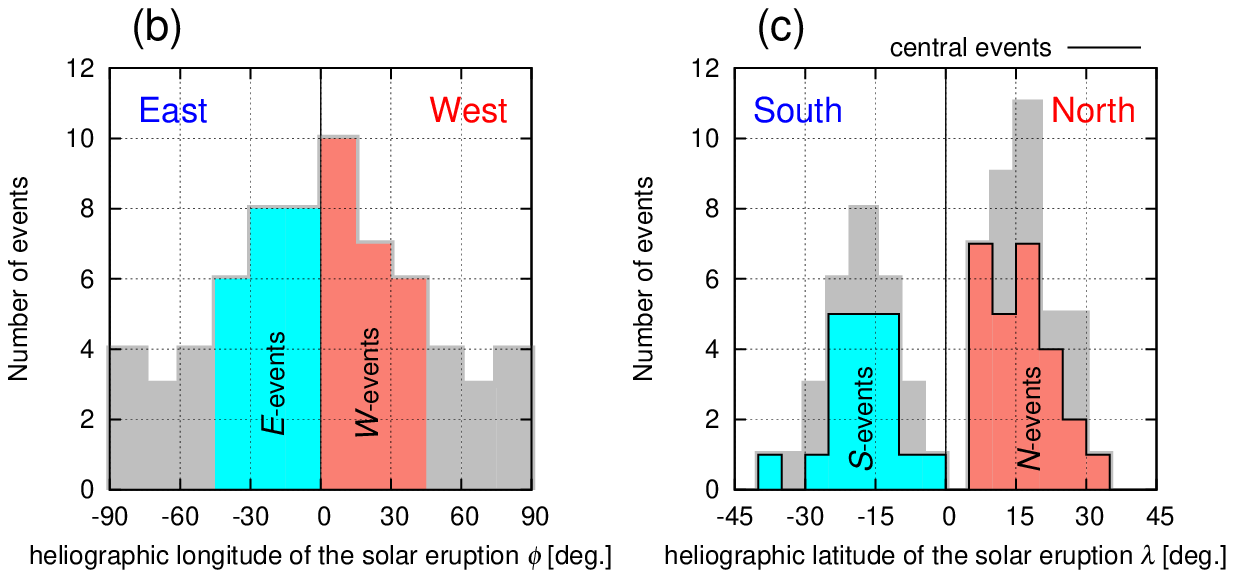}
   \caption{Heliographic locations of solar flares and filament disappearances associated with the 67 events in Table \ref{tb:ev}.
   The top panel displays the locations plotted on the solar surface (a),
while the bottom panels display histograms of the heliographic longitude, $\phi$ (b) and latitude, $\lambda$ (c).
Black solid points in panel (a) indicate the ``central events'' in $-45^\circ \le \phi \le +45^\circ$.
A red number in panel (a) indicates the event number in each region on the sun divided by black solid lines,
equator line ($\lambda=0^\circ$) and 5 meridian lines ($\phi=-90^\circ,-45^\circ,0^\circ,+45^\circ,+90^\circ$).
Blue and red histograms in panel (b) represent the $E$- and $W$-events
while those in panel (c) are the $S$- and $N$-events, groups in the central events.
For the definition of the $E$-, $W$-, $N$-, and $S$-events, see the text.}
   \label{loc}
\end{figure}

Table \ref{tb:ev} lists 79 CME events collected in this manner.
All the SSC onsets in the CME events coincide with discontinuous increases in solar wind speed,
magnetic field magnitude or proton density in the ACE or WIND data,
ensuring that the SSC can be used as an indicator of the IP-shock arrival in the CME event.
Solar event associations of 26 events in this table are also included in
the Richardson/Cane Near-Earth Interplanetary CMEs list\footnote{http://www.srl.caltech.edu/ACE/ASC/DATA/level3/icmetable2.htm} \citep{Can03,Ric10}.
From further analysis in this paper, we exclude 12 events noted with $\dagger$ or $\ddagger$ in Table \ref{tb:ev},
which lack the GMDN data or the location of the CME eruption in the SWPC data, and use the remaining 67 events.
Figure \ref{loc}a displays heliographic locations of the 67 CME eruptions on the solar surface.
Each red number in this figure indicates a number of CME eruptions in each heliographic region on the sun enclosed by
solid lines denoting the equator ($\lambda = 0^\circ$) and 5 meridians ($\phi = -90^\circ, -45^\circ, 0^\circ, 45^\circ, +90^\circ$).
The distribution of CME eruptions spreads over a wide range of longitude ($\phi$) as shown by the gray filled histogram in Figure \ref{loc}b,
allowing us to analyze the longitudinal distribution of GCRs behind the IP-shock.
It is also seen in Figure \ref{loc}b that the maximum number of events occurs around the longitudinal center as reported in previous studies \citep[e.g.][]{Gop07}.
The latitudinal ($\lambda$) distribution of the CME eruptions is, on the other hand, limited to the low- and mid-latitude zones
between 0$^\circ$-40$^\circ$ above and below the heliographic equator,
as shown by the gray filled histogram in Figure \ref{loc}c.\par
Out of the 67 CME events, we use for our superposition analyses only 45 events
associated with CME eruptions in the central region ($-45^\circ \le \phi \le +45^\circ$) on the sun (we call these events as ``central events''),
because the other 22 events associated with CME eruptions outside this region are known to show different properties when observed at Earth \citep{Gop07}.
In subsections \ref{sec:ssDen} and \ref{sec:ssEW}, we will perform superposition analyses for 22 ``$E$-events'' and 23 ``$W$-events''
of the central events associated with CME eruptions in eastern ($-45^\circ \le \phi < 0^\circ$) and western ($0^\circ \le \phi \le +45^\circ$) regions on the sun, respectively.
Blue and red histograms in Figure \ref{loc}b represent $\phi$ distributions in the $E$- and $W$-events.
In subsection \ref{sec:ssNS}, we will classify the central events into 26 ``$N$-events'' associated with northern ($\lambda > 0^\circ$) CME eruptions
and 19 ``$S$-events'' associated with southern ($\lambda < 0^\circ$) CME eruptions,
as represented by red and blue histograms in Figure \ref{loc}c.
                        
\section{Results}\label{sec:Res}
\subsection{Event samples}\label{sec:Samp}
We first present some event samples in this Section,
before we analyze the average spatial distribution of GCRs by superposing events.
Out of 45 events analyzed in this paper,
   we choose these events as samples in which
   (1) maximum depression in GCR density observed with NMs exceeds 3 \%,
   (2) there is no succeeding SSC onset within 2 days after the SSC onset
   under consideration, and
   (3) there is no deficit in the GMDN and NM data during the displayed time interval.

\subsubsection{2006 December 14 SSC event}\label{sec:ev1}
This SSC event is followed by a record intense geomagnetic storm with the maximum Kp index of +8.
The associated CME occurred following an X3.4 solar flare on December 13, 02:34 UT at S06W24.
A comprehensive view of this event is presented by \citet{Liu08} based on spacecraft data,
while \citet{Fus10b} analyzed a precursory ``loss-cone'' anisotropy observed with the GMDN prior to this event recorded at Earth.
We focus on the GCR density distribution observed after the SSC in the present paper.\par
\begin{figure}
   \epsscale{0.8}
   \plotone{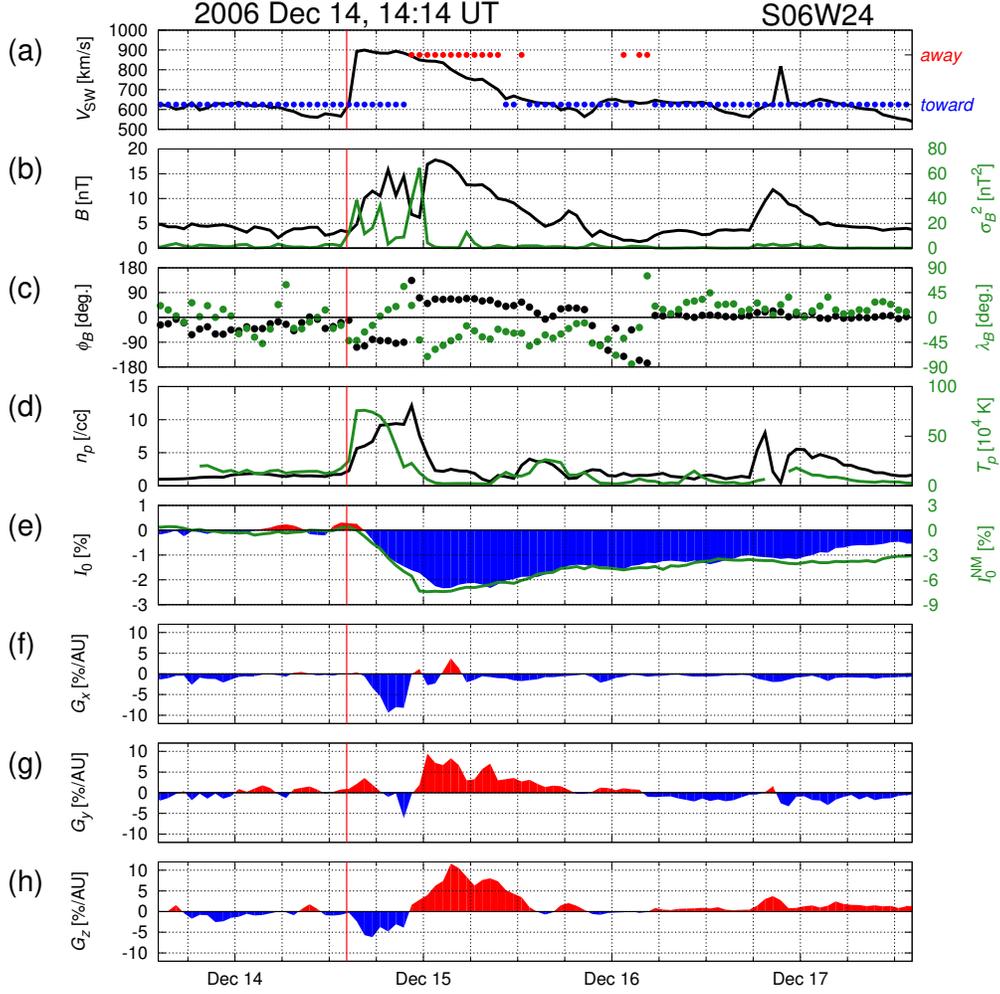}
   \caption{\footnotesize{A sample event following the SSC on 2006 December 14 at 14:14 UT.
      The heliographic location of the solar eruption associated with this SSC event is
      indicated above the right top corner of this figure.
      Panels from the top to the bottom display (a) hourly values of the solar wind speed ($V_{\rm SW}$),
      (b) magnetic field magnitude ($B$) and variance ($\sigma_B^2$),
      (c) GSE longitude ($\phi_B$) and latitude ($\lambda_B$) of the magnetic field orientation,
      (d) solar wind proton density ($n_p$) and temperature ($T_p$),
      (e) GCR density ($I_0$), and (f-h) GSE-$x, y, z$ components of the GCR density gradient (${\bm G}$),
      each as a function of time on the horizontal axis.
      The solar wind parameters in panels (a-d) are measured by ACE or WIND spacecraft.
      The GCR parameters in panels (e-h) are derived from the GMDN data, except for
      the green curve in panel (e) which is derived from NM data and shown on the right vertical axis.
      In panels (a-d), black and green curves or points are shown on the left and right vertical axes, respectively.
      Also the {\it away} and {\it toward} IMF sector polarities in each hour are respectively indicated
      by red and blue points in panel (a).
   The vertical red line in each panel indicates the SSC onset time.}}
   \label{ev1}
\end{figure}

Figure \ref{ev1} displays temporal variations of the solar wind data in panels (a) to (d),
the GCR density $I_0$ observed with the GMDN (color shaded curve) and NMs (green curve) in panel (e) and
three GSE components of the density gradient ${\bm G}$ derived from the GMDN data in panels (f) to (h),
all during the time interval from 1 day before the SSC onset to 3 days after the SSC onset.
The IMF sector polarity indicated by red and blue points in Figure \ref{ev1}a is designated
referring to the hourly mean magnetic field ${\bm B}(t)$ observed in the GSE coordinate system,
as {\it away} when $B_x<B_y$ and {\it toward} when $B_x>B_y$, as expected from the Parker's spiral magnetic field.
The variance of the magnetic field, $\sigma_B^2(t)$ displayed by a green curve in Figure \ref{ev1}b is derived on an hourly basis as
\begin{equation}
   \sigma_B^2(t) = \frac{1}{3\times60} \sum_{i=1}^{60} \left\{ \left(b_x^i(t) - B_x(t)\right)^2 + \left(b_y^i(t) - B_y(t)\right)^2 + \left(b_z^i(t) - B_z(t)\right)^2 \right\}
\end{equation}
where ${\bm b}^i(t)$ ($i=1,2,\cdots,60$) is a minute average of the magnetic field in a temporal interval $t \sim t+1$ hours.
The GCR densities, $I_0(t)$ and $I_0^{\rm NM}(t)$ are normalized to the 6 hours average prior to the SSC onset.\par
As reported by \citet{Liu08}, the azimuthal angle $\phi_B$ of the magnetic field orientation in Figure \ref{ev1}c
shows a monotonic rotation during one day after the end of December 14, indicating a Magnetic Flux Rope (MFR) passing Earth.
The $G_x$ in Figure \ref{ev1}f shows a negative enhancement after the SSC onset until the end of the magnetic sheath region behind IP-shock,
corresponding to the decreasing phase of the density in Figure \ref{ev1}e.
This is consistent with a density minimum approaching Earth from the sunward direction ($x>0$)
and being observed as a negative enhancement of $G_x$.
Following the sheath region, positive $G_y$ and $G_z$ in Figures \ref{ev1}g and \ref{ev1}h are
clearly enhanced when Earth enters the minimum density region inside the MFR,
indicating that the density minimum passed the south-west of Earth ($y<0$ and $z<0$)
after propagating radially outward from the CME eruption on the sun.
According to \citet{Liu08}, the GSE latitude and longitude of the MFR axis orientation
best-fitted to the spacecraft data are $\sim 60^\circ$
and $\sim 270^\circ$ in the GSE-coordinate, respectively, and the axis passed the west of Earth.
The density gradient in Figure \ref{ev1} is consistent with the GCR density minimum located
on the MFR axis approaching and leaving Earth.
\citet{Kuw04,Kuw09} analyzed the density gradient vector derived from the GMDN data and
deduced the cylinder geometry of the GCR depleted region in CMEs.
The next SSC is also recorded on 2006 Dec. 16 (see Table \ref{tb:ev}) within the time interval displayed in Figure \ref{ev1} and
is associated with the CME following an X1.5 solar flare at S06W46.\par
During the first event, the GCR density, $I_0(t)$, derived from the GMDN data in Figure \ref{ev1}e shows
a similar variation to $I_0^{\rm NM}(t)$ (green curve) derived from NM data
which is free from the atmospheric temperature effect.
This implies that the GCR density is properly derived from the GMDN data in this event by our analysis method,
even though the temporal variation of hourly $I_0(t)$ may potentially include
$\sim 0.18$ \% influence from the temperature effect as mentioned in Section \ref{sec:AnaXi}.
We note that the magnitude of the FD is larger in $I_0^{\rm NM}$ derived from NM data than in $I_0$ derived from the GMDN data,
indicating a soft rigidity spectrum of the density depression in the FD.

\subsubsection{2012 June 16 SSC event}\label{sec:ev2}
\begin{figure}
   \epsscale{1}
      \plotone{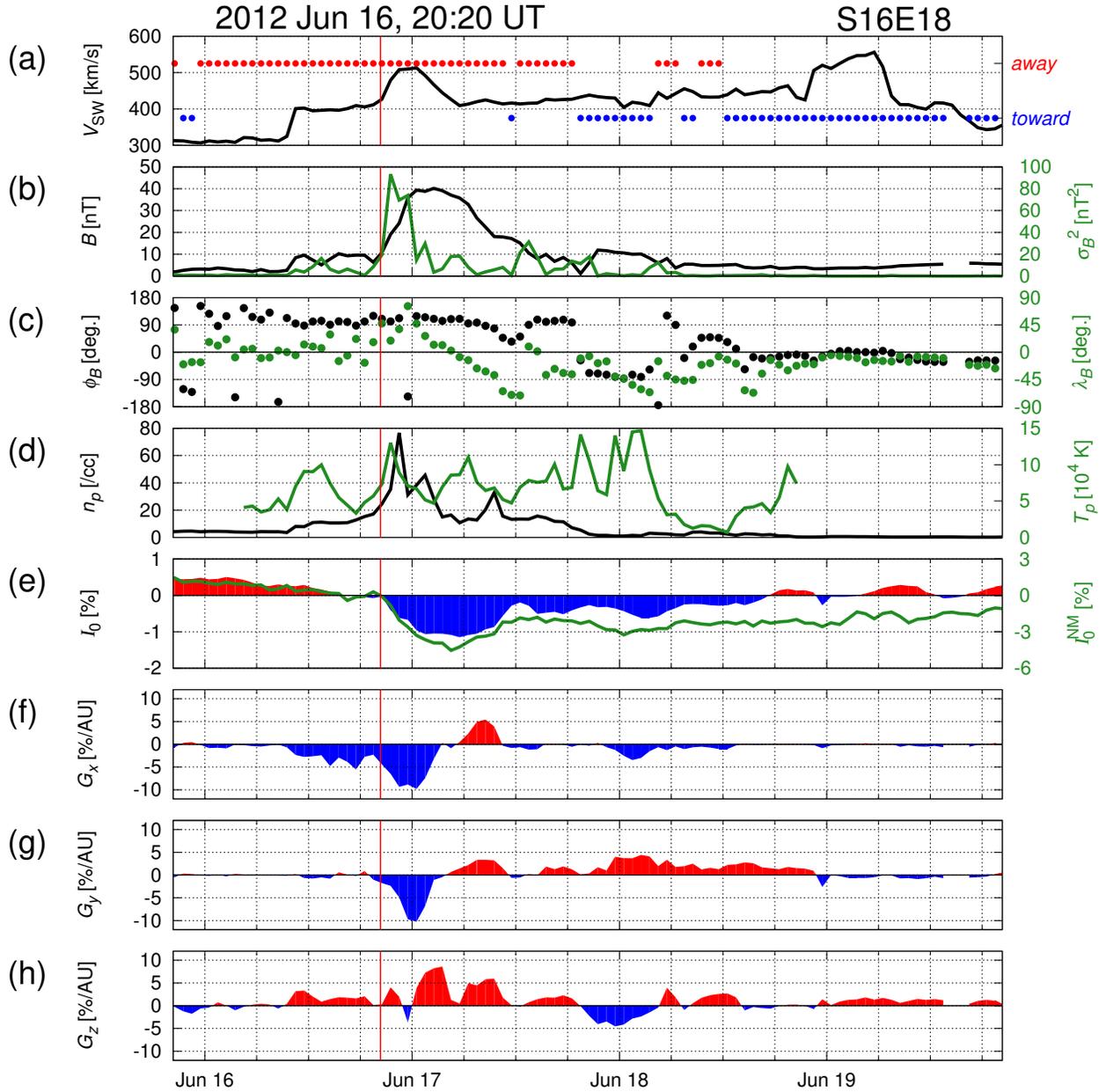}
      \caption{A sample event following the SSC on 2012 June 16 at 20:20 UT displayed in the same manner as Figure \ref{ev1}.}
   \label{ev2}
\end{figure}

This SSC event, displayed in Figure \ref{ev2}, is associated with a CME which erupted from the sun accompanying an M1.2 solar flare on June 13, 13:41 UT at S16E18.
The $G_y$ and $G_z$ in Figures \ref{ev2}g and \ref{ev2}h show negative and positive enhancements, respectively,
indicating that the density minimum region passed the south-east of Earth after propagating radially outward from the CME eruption on the sun.
A nearly $180^\circ$ rotation of the magnetic field latitude $\lambda_B$ in Figure \ref{ev2}c accompanied by the rapid decrease and recovery of $I_0$ in Figure \ref{ev2}e
indicates a MFR passing Earth in the first half of June 17.
During the same period, ecliptic components of the gradient, $G_x$ and $G_y$
in Figures \ref{ev2}f and \ref{ev2}g,
show clear reversals from negative to positive when Earth passes near the density minimum in the MFR.
The $G_z$ remains positive during the same period possibly indicating the density minimum passed the south of Earth.
It should be noted, however, that Earth is mostly in the {\it away}
IMF sector during this period as indicated by red points in Figure \ref{ev2}a
and the positive $G_z$ is also expected from the drift model for
the large-scale GCR transportation in the {\it away} sector.
The positive $G_z$ in the 2006 December 14 SSC event is also observed mostly
in the {\it away} sector (see Figures \ref{ev1}a and \ref{ev1}h).
We will analyze this effect in detail later in Section \ref{sec:ssNS}.\par
We note again that the overall temporal variations of $I_0$ and $I_0^{\rm NM}$ in Figure \ref{ev2}e are similar to each other,
while the magnitude of the density depression in the FD is significantly larger in $I_0^{\rm NM}$ than in $I_0$,
indicating a soft rigidity spectrum of the density depression in the FD.
The depression in $I_0$ finishes by the end of June 18 while that in $I_0^{\rm NM}$ lasts over June 19,
possibly indicating the rigidity dependence of the recovery from the density depression,
i.e. the density depression of higher rigidity GCRs recoveries faster.
We can also see, however, that the solar wind velocity $V_{\rm SW}$ in Figure \ref{ev2}a is enhanced again at the end of June 18,
which may possibly cause the long duration of $I_0^{\rm NM}$ depression if this affects more effectively on $I_0^{\rm NM}$ than on $I_0$.
This enhancement of $V_{\rm SW}$ is not considered as an IP-shock event,
because there is no enhancements seen at the same time in other solar wind parameters shown in Figures \ref{ev2}b and d.

\subsubsection{2013 April 13 SSC event}
\begin{figure}
   \plotone{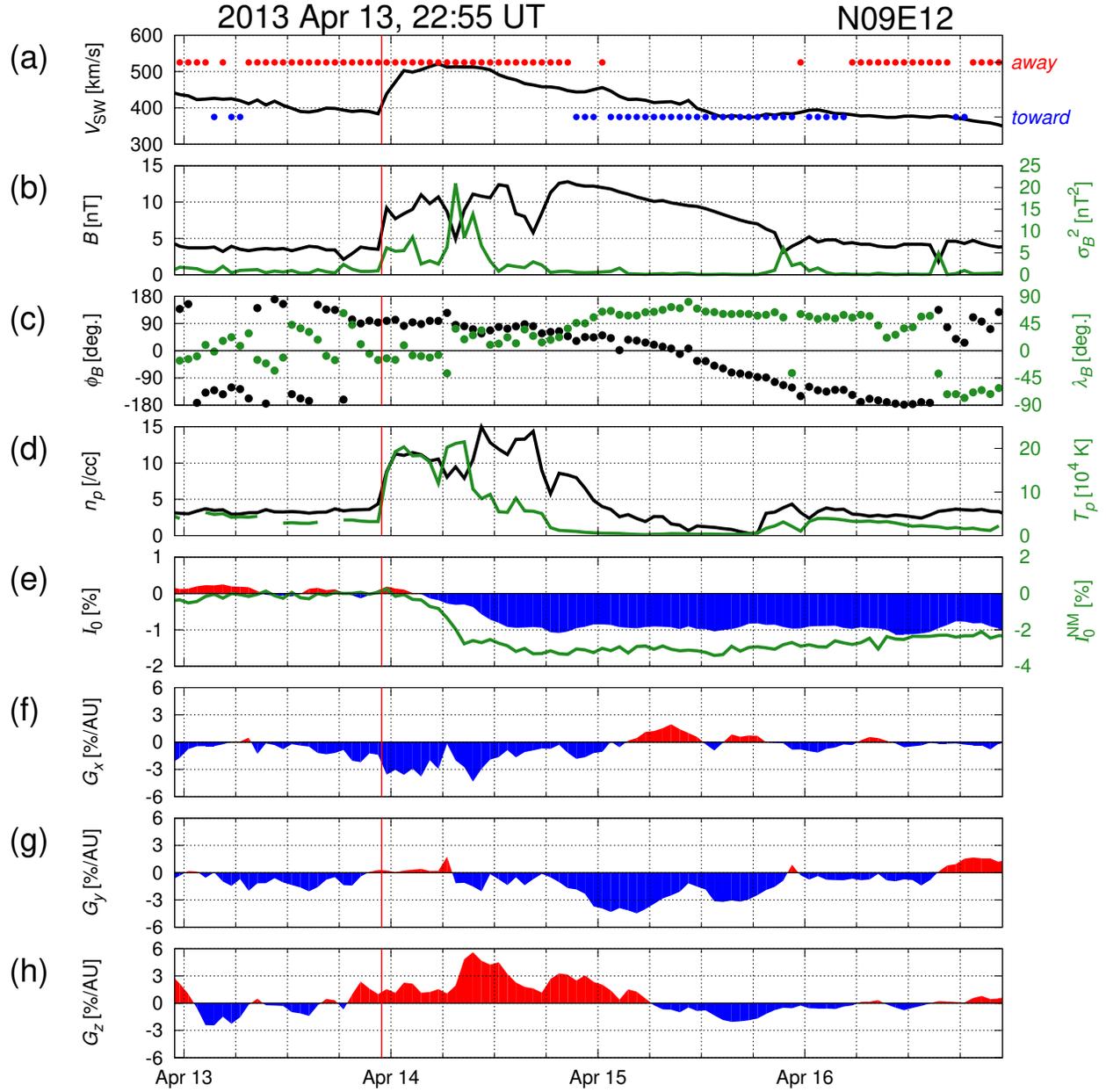}
      \caption{A sample event following the SSC on 2013 April 13 at 22:55 UT displayed in the same manner as Figure \ref{ev1}.}
   \label{ev3}
\end{figure}

This SSC event, displayed in Figure \ref{ev3}, is associated with a CME which erupted from the sun accompanying an M6.5 solar flare on April 11, 07:10 UT at N09E12.
A monotonic rotation of $\phi_B$ in Figure \ref{ev3}c and decreases of the proton density $n_p$ and temperature $T_p$ in Figure \ref{ev3}d
indicate an MFR passing Earth during a day after 18:00 on April 14,
but it shows only a minor effect on the GCR density $I_0$ and $I_0^{\rm NM}$ in Figure \ref{ev3}e.
The $G_y$ in Figure \ref{ev3}g, on the other hand, shows a negative enhancement during the MFR period
in accord with the GCR density minimum region propagating radially outward from the CME eruption on the sun.
The $G_z$ shows a clear reversal of its sign from positive to negative during the MFR period.
The $G_x$ and $G_y$ in 2012 June 16 SSC event displayed in the previous subsection also showed similar reversals.
This typically demonstrates an advantage of the density gradient (or anisotropy) observations
in deriving a three-dimensional geometry of the GCR depleted region in the MFR,
while it is difficult to deduce that only from the observed GCR density ($I_0$ and $I_0^{\rm NM}$).

\subsection{Superposition analysis and the average spatial distribution of GCR density}\label{sec:ss}
In this section, we perform a superposition analysis of the 45 central events and deduce the average spatial distribution of GCRs.
As seen in sample events in Section \ref{sec:Samp}, all events show different temporal profiles of the solar wind parameters,
i.e. the duration and magnitude of the solar wind and magnetic field enhancements, the duration of the magnetic sheath and the MFR signatures following the sheath,
are all different between one event and the next, causing different temporal variations in $I_0$ and ${\bm G}$.
We cannot derive these individual features of each event from the superposition analysis which simply averages out these features.
Analyses of the GMDN data for deriving individual event features can be found elsewhere \citep{Mun03,Kuw04,Kuw09}.
The superposition analysis allows us to discuss the average features of $I_0$ and ${\bm G}$ which reflect the average spatial distribution of GCRs behind IP-shock.
This is our motivation of the superposition analyses presented below.

\subsubsection{Conversion of temporal variations to spatial distributions of the GCR density and gradient}\label{sec:conv}
The temporal variations of the solar wind parameters and the GCR density and density gradient analyzed in the preceding sections represent
spatial distributions of those parameters convected radially outward by the solar wind and observed at fixed locations on Earth.
Due to the difference in the average solar wind velocity, however, even an identical spatial distribution may result in different temporal variations.
In order to deduce average spatial distributions more accurately from the superposition analysis presented in the following subsections,
we first convert the temporal variations to the spatial distributions.
By assuming the spatial distribution of a quantity $Q(x,y,z)$ in steady state in the solar wind frame,
the temporal variation of $Q$ ($Q_E(t)$) at Earth ($x_E=0, y_E=0, z_E=0$) is related to the spatial distribution of $Q$, as
\begin{equation}
   Q_E(t) = Q(V_{\rm SW}(t)t,0,0)
\end{equation}
where $t$ is the time measured from the SSC onset at $t=0$ and
$V_{\rm SW}(t)$ is the solar wind velocity measured at Earth at $t$.
Thus, the time $t$ can be converted to the GSE coordinate $x$ as
\begin{equation}
   x = V_{\rm SW}(t) t.
   \label{eq:vt}
\end{equation}\par
It is noted, however, that the conversion by equation (\ref{eq:vt}) may cause the following technical problem.
According to equation (\ref{eq:vt}), two separate times $t_1$ and $t_2$ correspond to $x_1 = V_{\rm SW}(t_1) t_1$ and $x_2 = V_{\rm SW}(t_2) t_2$, respectively,
and, in case of $V_{\rm SW}(t_1) \leq V_{\rm SW}(t_2)$, we can keep $t$ and $x$ in the same order, i.e. $x_1<x_2$ if $t_1<t_2$.
In case of $V_{\rm SW}(t_1) > V_{\rm SW}(t_2)$, on the other hand, we may get $x_1>x_2$ even if $t_1<t_2$.
To avoid this problem and keep $x$ and $t$ in the same order, we make the conversion, as
\begin{equation}
   x(t) =\sum_{k=0}^{t/\Delta t} V_{\rm SW}(k \Delta t) \Delta t
   \label{eq:trans}
\end{equation}
where $k$ is the time in units of $\Delta t$ and $\Delta t$ is set at $\Delta t = 1$ hour
corresponding to the hourly count rate analyzed in this paper.
Note that $x>0$ ($x<0$) corresponds to $t>0$ ($t<0$) after (before) the SSC onset and $x$ increases
toward the sun (i.e. GSE-$x$ direction) with $x=0$ corresponding to the IP-shock arrival at Earth at $t=0$.
The $x = 0.1$ AU roughly corresponds to $t \sim 9$ hours in the case of $V_{\rm SW} = 450$ km/s.\par
The $x$ calculated by equation (\ref{eq:trans}) may not give us a real spatial coordinate,
because we assume that the spatial distribution of GCRs is constant
   on the solar wind frame
   propagating radially outward with solar wind velocity $V_{\rm SW}(t)$
   at $t$ at Earth.
The real spatial distribution actually varies even on the solar wind frame due to, for instance,
the expansion of the CME during the propagation past Earth.
Even so, the conversion gives us an estimation of the spatial scale of the GCR distribution
in the FD in the vicinity of Earth that is the main subject of the present paper.
Moreover, the conversion also works for correcting each event for the difference in the average solar wind speed.
It is noted that we confirmed all conclusions derived in this paper remaining essentially unchanged before and after the conversion.

\subsubsection{Average features of the GCR density distribution}\label{sec:ssDen}
\begin{figure}
   \epsscale{0.95}
   \plotone{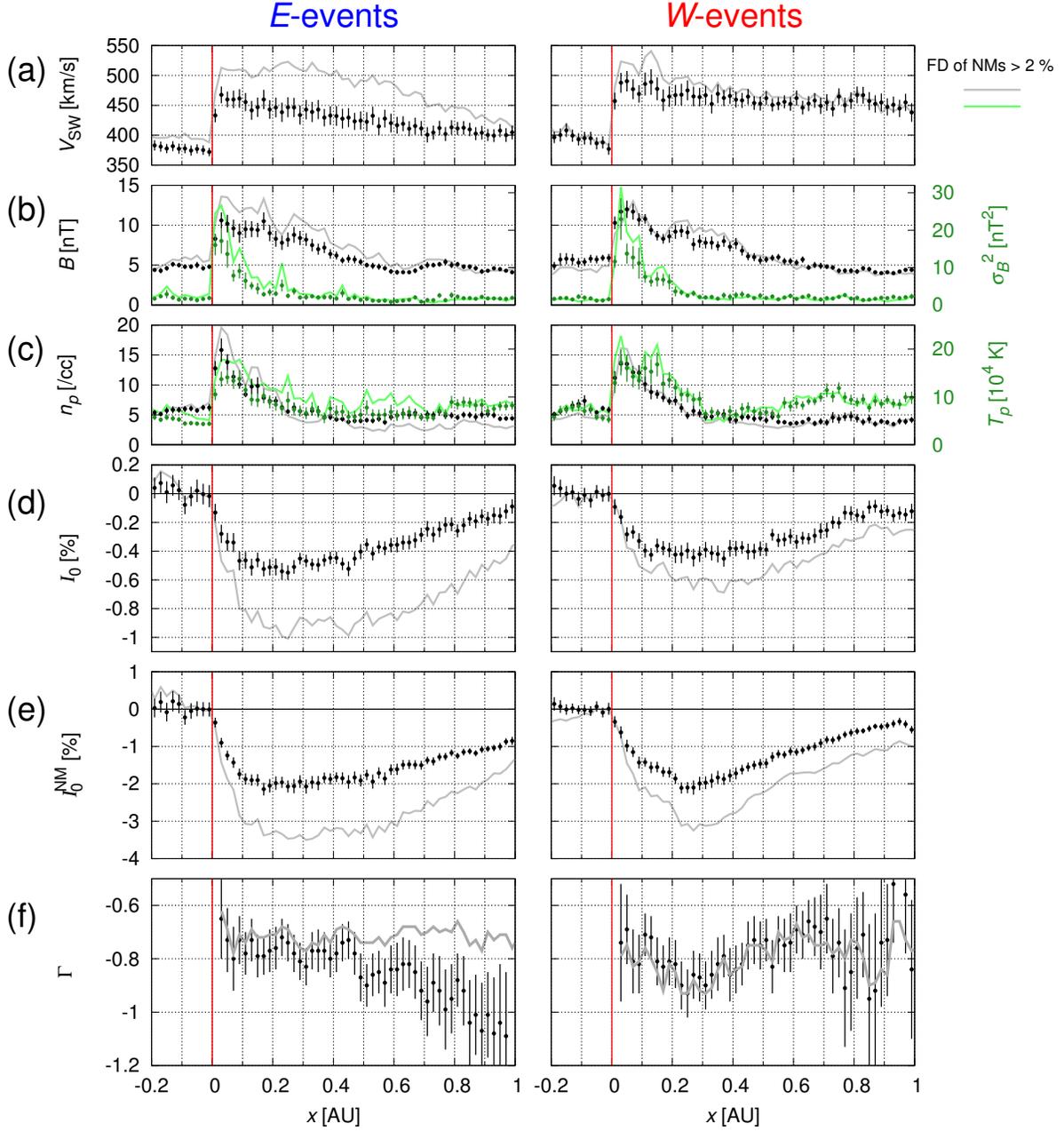}
   \caption{\scriptsize{Averages of the superposed spatial distribution of the solar wind parameters and GCR density:
      (a) solar wind speed ($V_{\rm SW}$),
      (b) magnetic field magnitude ($B$) and variance ($\sigma_B^2$) measured by the ACE or WIND spacecraft,
      (c) solar wind proton density ($n_p$) and temperature ($T_p$),
      (d) GCR density ($I_0$) observed with the GMDN,
      (e) GCR density ($I_0^{\rm NM}$) observed with NMs, and
      (f) exponent ($\Gamma$) of the power-law rigidity spectrum of the GCR density depression,
      each as a function of GSE-$x$ in AU measured from the SSC onset at $x=0$ (or $t=0$ in time $t$) indicated by a vertical red line.
      For the conversion from recorded time $t$ to GSE-$x$, see the text.
      Black (green) point and error in panels (a-c), each shown on the left (right) vertical axis,
      are derived from the average and dispersion of the superposed distributions in every $\Delta x=0.02$ AU on the horizontal axis.
      In panel (f), a black point is derived from the black points in panels (d) and (e) (see text),
      while an erorr bar is evaluated by an error propagation from errors in panels (d) and (e).
      Gray and green curves in each panel display the averages of the intense events in which the maximum density depressions in FDs derived from NM data exceed 2 \%,
      each shown on the left and right vertical axes, respectively.
      Left panels display the $E$-events, while right panels display the $W$-events (see Figure \ref{loc} and text).}}
   \label{ssI}
\end{figure}

Figure \ref{ssI} shows the superposed spatial distributions of the solar wind speed ($V_{\rm SW}$), IMF magnitude ($B$) and variance ($\sigma_B^2$),
proton density ($n_p$) and temperature ($T_p$), GCR densities derived from the GMDN and NM data ($I_0$ and $I_0^{\rm NM}$),
and exponent ($\Gamma$) of the power-law rigidity spectrum of the density depression estimated from $I_0$ and $I_0^{\rm NM}$,
each as a function of GSE-$x$ in AU on the horizontal axis which is calculated from equation (\ref{eq:trans}).
The left and right panels display the superpositions of the $E$- and  $W$-events defined in Section \ref{sec:IdEv}, respectively.
In each panel, a black (green) point and error on the left (right) vertical axis are derived from the average and
dispersion of the superposed spatial distributions in every $\Delta x = 0.02$ AU on the horizontal axis.
The gray (green) curve shown on the left (right) vertical axis is the average of the intense events
in which the maximum density depression derived from NM data exceeds 2 \% (see Table \ref{tb:ev}).
A range of -0.2 AU $<x<$ +1 AU is covered in this figure. In the case of more than two IP-shocks being recorded within this range,
we use only the data before (after) the following (previous) SSC onset for the superposition throughout this paper,
in order to minimize the interference between separate events without losing events from our analysis.\par
In order to remove longer-term density variations superposed
   on the short-term decreases in FDs,
   we normalize the densities ($I_0(t)$ and $I_0^{\rm NM}(t)$)
   to the averages over a 5-day period
   beginning one day prior to the SSC onset in each SSC event.
   We calculate a deviation of the density at each time $t$
   from the 5-day average in each event and
derive an error of the average density in each $\Delta x$ bin
from the dispersion of the deviations in all events analyzed.
After the superposition,
average spatial distributions of $I_0(x)$ (Figure \ref{ssI}d)
and $I_0^{\rm NM}(x)$ (Figure \ref{ssI}e) are normalized to the averages
over 0.06 AU in -0.06 AU $<x<$ 0 AU.
Each of them generally shows an abrupt decrease at $x=0$ AU followed by a gradual recovery continuing up to $x=+1$ AU,
i.e. a well-known feature of typical FDs. Looking at this figure closer, we also find that the initial decreasing phase of $I_0(x)$ and $I_0^{\rm NM}(x)$
(the left panels of Figures \ref{ssI}d and \ref{ssI}e) in the E-events ends within 0 AU $<x<+0.1$ AU,
the sheath region behind IP-shock as indicated by the enhanced $\sigma_B^2$, $n_p$, and $T_p$ in Figures \ref{ssI}b and \ref{ssI}c.
This is consistent with numerical calculations \citep[e.g.][]{Nis82} of the ``propagating diffusive barrier'' model mentioned in Section \ref{sec:Int},
indicating that cosmic ray modulation by the compressed magnetic field sheath is a main cause of the GCR density depression in the $E$-events.
The initial decreasing phase of $I_0(x)$ and $I_0^{\rm NM}(x)$ (the right panels of Figures \ref{ssI}d and \ref{ssI}e) in the $W$-events,
on the other hand, spreads wider beyond $x=+0.1$ AU with a slower decreasing rate than in the $E$-events.
Around the region of +0.1 AU $<x<$ +0.2 AU, depressions of $I_0$ and $I_0^{\rm NM}$ in $E$-events are deeper than
those in $W$-events.
During $W$-events Earth encounters the eastern flank of the IP-shock.
The slower decrease of GCR density in the magnetic sheath in such events can be attributed to a weaker compression of the IMF
in the eastern flank of the IP-shock, as discussed in Section \ref{sec:Int} \citep{Can94}.
The wider sheath region in the $W$-events can be actually seen in $T_p$ in the right panel of Figure \ref{ssI}c.
This E-W asymmetry of GCR modulation in the sheath region is seen more clearly in intense events displayed by
gray curves in Figures \ref{ssI}d and \ref{ssI}e.\par
After the initial decreasing phase, $I_0(x)$ and $I_0^{\rm NM}(x)$ also show broad minima followed by gradual recoveries.
This is due to an additional GCR modulation in the central CME region (or CME ejecta) behind the sheath region,
which is typically indicated by a broad pit of $T_p$ in the right panel of Figure \ref{ssI}c.
The magnetic flux rope (MFR) often formed in the CME ejecta excludes GCRs from penetrating into the MFR by
its adiabatic expansion, sometimes causing prominent GCR decreases.
The GCR density depression in FDs is generally caused by these two distinct modulations, respectively in the sheath and central CME regions.
The modulation in the central CME region is seen clearer in $I_0(x)$ and $I_0^{\rm NM}(x)$ in $W$-events
(the right panels of Figures \ref{ssI}d and \ref{ssI}e) than in $E$-events in the left panels,
because of the weaker modulation in the sheath region due to the E-W asymmetry mentioned above.
The modulation is also seen in intense $E$-events displayed by gray curves in the left panels as broad minima extending over +0.1 AU $<x<+0.5$ AU,
but the density depression is much larger in the sheath region.
The maximum depression of GCR density $I_0(x)$ by the GMDN (black points in Figure \ref{ssI}d) is slightly larger in $E$-events than in $W$-events
in accord with the E-W asymmetry in the FD magnitude mentioned in Section \ref{sec:Int},
while the asymmetry is clearer in intense events (gray curves in Figure \ref{ssI}d).
This is probably due to the larger E-W asymmetry of the GCR modulation in the sheath in intense IP-shocks.
If we look at $I_0^{\rm NM}(x)$ by NMs in Figure \ref{ssI}e), however,
no such clear E-W asymmetry is seen in the maximum depression even in the intense events.
This is because the relative contribution of the modulation in the central CME region to the total GCR modulation is
larger in $\sim 10$ GV GCRs monitored by NMs than in $\sim 60$ GV GCRs observed by the GMDN.\par
By comparing the average density distributions in E- and W-events
   in Figure \ref{ssI}d (black points with error),
   we find that the difference between $I_0(x)$s in $E$- and $W$-events is
   only one or two times the error
and the statistical significance of the difference at each $x$ is not high.
The difference (one above/below the other), however, continues over
successive $\sim 10$ $x$-values in the same sense,
indicating that the systematic difference is statistically significant.
As discussed in Section \ref{sec:AnaXi}, we also obtained $\sim 0.18$ \% as a measure of the influence to hourly $I_0(t)$
from the atmospheric temperature effect.
The standard error due to the temperature effect of the average of $\sim 20$ events superposed, therefore,
is estimated to be $0.18/\sqrt{20-1}=0.04$ \%. Similar error is expected from the temperature effect for each data point in  Figure \ref{ssI}d,
because each $x$ bin with a width of $\Delta x=0.02$ AU corresponds to
$\sim 2$ hours in the time-to-space conversion in Section \ref{sec:conv} with $V_{\rm SW}=450$ km/s
and contains 1 or 2 hourly data in each event.
This error of $0.04$ \% is less than the error bars in Figure \ref{ssI}d,
indicating that the temperature effect does not alter results described above.\par
The rigidity dependence of GCR density depression can be quantitatively evaluated from the comparison between $I_0(x)$ and $I_0^{\rm NM}(x)$
in Figures \ref{ssI}d and \ref{ssI}e. On an assumption of a power-law dependence ($P^\Gamma$) of the density depression on the GCR rigidity ($P$),
the power-law index $\Gamma$ can be given by the ratio $I_0(x)/I_0^{\rm NM}(x)$ as
\begin{equation}
   \Gamma(x) = \frac{\ln\left( I_0(x)/I_0^{\rm NM}(x) \right)}{\ln\left( P_{\rm GMDN}/P_{\rm NM} \right)}
\end{equation}
where $P_{\rm GMDN} = 60$ GV and $P_{\rm NM} = 10$ GV are representative median rigidities of primary GCRs observed with the GMDN and NMs, respectively.
Figure \ref{ssI}f displays $\Gamma(x)$ as a function of $x$. The black points in Figure \ref{ssI}f indicate $\Gamma$ derived from
the black points in Figures \ref{ssI}d and \ref{ssI}e, while the gray curve in Figure \ref{ssI}f shows $\Gamma$ derived from
the gray curves in Figures \ref{ssI}d and \ref{ssI}e for intense FDs.
It is seen that $\Gamma$ varies in a range of -1.2 $<\Gamma<$ -0.6 in accord with
most of the previous studies reporting $\Gamma \sim -1$ \citep{Loc60,Wad80,Sak85,Sak87,Mor90}.
The black points in $E$-events in the left panel of Figure \ref{ssI}f show a rapid decrease with increasing $x$
during the recovery phase of the FD in $x>+0.4$ AU, implying that higher rigidity (60 GV) GCRs recover faster than lower rigidity (10 GV) GCRs.
The $\Gamma$ in intense $E$-events displayed by a gray curve in Figure \ref{ssI}f,
on the other hand, shows no such rapid decrease in $E$-events, remaining at $\Gamma \sim -0.7$ up to $x=+1$ AU.
This is due to the faster and stronger shocks, as indicated by gray and green curves in Figures \ref{ssI}a-\ref{ssI}c,
preventing even high energy GCRs from refilling the density depleted region in FDs.
The $\Gamma$ in $W$-events (black points in the right panel of Figure \ref{ssI}f) also shows no rapid decrease,
probably due to the longer duration of the enhanced solar wind velocity as shown in the right panel of Figure \ref{ssI}a,
which is similar to the gray curve in the left panel. It is interesting to note that the $\Gamma$ in $W$-events shows
a transit decrease to $\Gamma \sim -0.9$ in +0.2 AU $<x<+0.4$ AU where $T_p$ in the right panel of Figure \ref{ssI}c decreases
and the modulation in the central CME region is observed in the right panels of Figures \ref{ssI}d and \ref{ssI}e.
Due to this transit decrease, $\Gamma$ in $W$-events is smaller in $E$-events at $x \sim$ +0.25 AU.
Due to a large error, this difference between $\Gamma$s in  $E$ and $W$-events at each $x$ is only one or two sigma.
The difference, however, again continues over successive $\sim 10$ $x$-values in the same sense,
indicating that the systematic difference is statistically significant.
This implies that the modulation in the central CME region
has a softer rigidity spectrum than the modulation in the sheath region.
Due to this rigidity dependence, the density depression in the central CME region dominates the total depression in FD in $I_0^{\rm NM}(x)$.
This is consistent with the E-W asymmetry of the maximum density depression due to the modulation in the sheath region
being seen only in $I_0(x)$ by the GMDN but not in $I_0^{\rm NM}(x)$ by NMs.

\subsubsection{GCR density gradient in the ecliptic plane}\label{sec:ssEW}
\begin{figure}
   \plotone{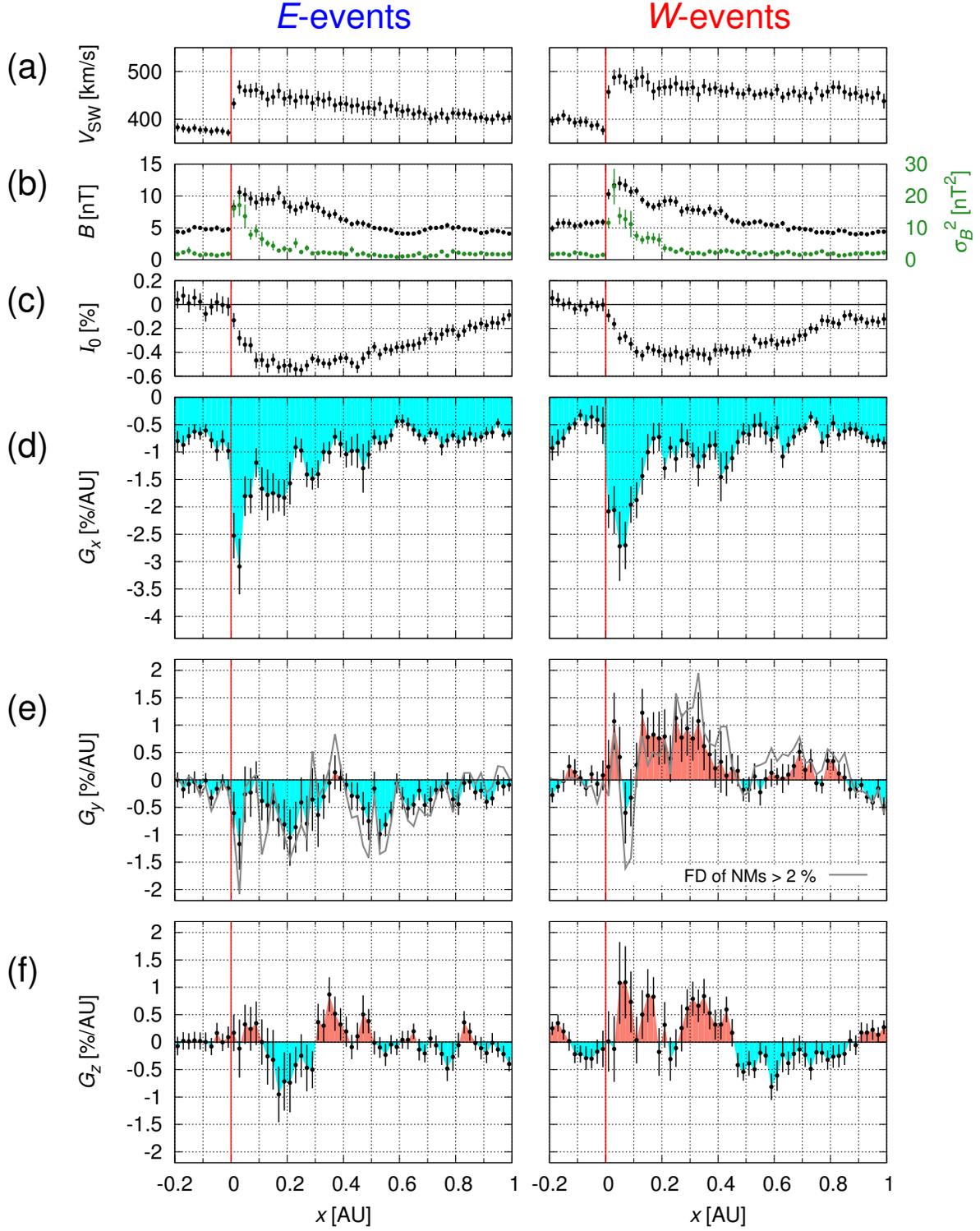}
   \caption{\footnotesize{Averages of the superposed spatial distributions:
      (a-c) solar wind speed ($V_{\rm SW}$), magnetic field magnitude ($B$) and variance ($\sigma_B^2$),
      and GCR density ($I_0$) in the same manner as Figures \ref{ssI}a, \ref{ssI}b, and \ref{ssI}d,
      and (d-f) three GSE components of the GCR density gradient ($G_x$, $G_y$, $G_z$) derived from the GMDN data.
      The format is the same as Figure \ref{ssI}.}}
   \label{ssGew}
\end{figure}

Figure \ref{ssGew} shows the superposed spatial distributions of
the solar wind parameters and the GCR density and gradient in the $E$- and $W$-events, in the same manner as Figure \ref{ssI}.
Before the SSC onset ($x<0$), the average $G_x$ in Figure \ref{ssGew}d has a negative offset of $\sim -1$ \%/AU due to
the radial density gradient in the steady state arising from the solar wind convection of the GCR particles \citep[cf.][]{Par65,Mun14}.
Following the SSC onset ($x>0$), the negative $G_x$ in Figure \ref{ssGew}d shows a clear enhancement immediately behind the IP-shock.
This enhancement extends $\sim 0.2$ AU in $W$-events, while it extends $\sim 0.1$ AU in $E$-events.
This E-W asymmetry of $G_x$ corresponds to the longer initial decreasing phase of the density $I_0$ (Figure \ref{ssGew}c)
in the $W$-events discussed in the previous subsection.
It is shown in Appendix \ref{sec:Gx} that $G_x$ in Figure \ref{ssGew}d is consistent with
the spatial derivative of $I_0$ in Figure \ref{ssGew}c ($dI_0(x)/dx$).\par
The average distribution of $G_y$ in Figure \ref{ssGew}e shows a broad negative (westward gradient) enhancement behind the IP-shock in $E$-events
while it shows a positive (eastward gradient) enhancement in $W$-events.
The eastward (westward) gradient on the east (west) side of the central CME implies
that the GCR density minimum is located around the longitudinal center behind the IP-shock,
in accord with the center-limb effect suggested by \citet{Yos65}.
This is also confirmed in the gray curve in Figure \ref{ssGew}e,
the superposition of the intense events in which the maximum density depressions
in FDs derived from NM data exceed 2 \%.\par
\begin{figure}
   \epsscale{1}
   \plotone{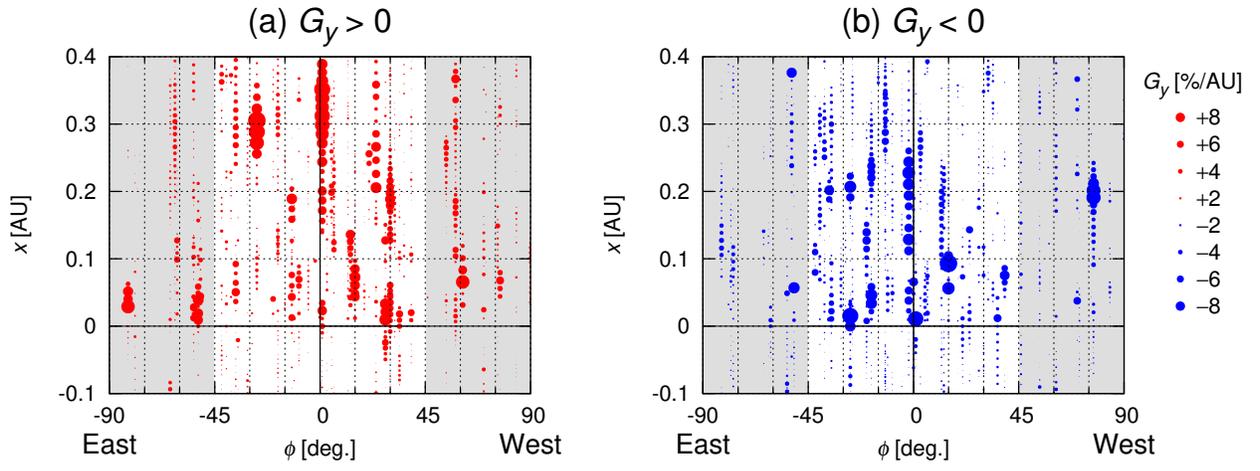}
   \caption{Spatial distribution of $G_y$ classified according to the value of $G_y$:
      (a) distribution of positive $G_y$ and (b) negative $G_y$.
      Different marks refer to different domains of $G_y$ (see right of panel (b)).
      Solid circles along a vertical line display $G_y$ in an event as a function of GSE-$x$ on the vertical axis,
   while the horizontal axis represents the heliographic longitude ($\phi$) of the solar eruption associated with each event.}
   \label{lon_gy}
\end{figure}

Figure \ref{lon_gy} shows ``bubble plots'' representing the spatial distribution of $G_y$ classified according to the value of $G_y$.
Different marks refer to different domains of $G_y$ (see right of panel (b)).
Panels (a) and (b) show positive and negative $G_y$ separately.
Solid circles plotted along a vertical line represent all $G_y$ observed during an event as a function of GSE-$x$ on the vertical axis
while the horizontal axis represents the heliographic longitude ($\phi$) of the location of the solar eruption associated with each event.
The shaded area represents the heliographic region ($\phi<-45^\circ$ and $\phi>+45^\circ$) outside the central region on the sun,
in which the CME events are excluded from the superposed epoch analysis.
The positive $G_y$ (red circles in panel (a)) is seen to be dominant in western ($\phi>0^\circ$) events
while negative $G_y$ (blue circles in panel (b)) is dominant in eastern ($\phi<0^\circ$) events.
This asymmetry results in the enhancements with opposite signs in Figure \ref{ssGew}e.
The spatial extent of the enhancement seems to be larger in $G_y$ than in $G_x$,
as seen in Figures \ref{ssGew}d and \ref{ssGew}e.
It is interesting to note that $G_y$ in Figure \ref{ssGew}e shows
simultaneous enhancements in 0 AU $<x<$ +0.1 AU with opposite signs in $E$- and $W$-events,
which are possibly related to the sheath structure between IP-shock and the CME ejecta.\par
The north-south component of the density gradient, $G_z$, in Figure \ref{ssGew}f
also shows a positive enhancement after the SSC onset, particularly in $W$-events,
but this can be attributed to a north-south asymmetry of the density depression in the FDs analyzed in this paper.
As shown in the next section, the $S$-events have a significantly deeper density depression than the $N$-events.
This implies that the GCR density minima propagating radially outward
from the CME eruptions on the sun were deeper when they passed south of Earth,
resulting in the positive $G_z$ (northward gradient) enhancement in the right panel of Figure \ref{ssGew}f.
This may be the case also in the $E$-events,
but the number of $E$-events is almost twice as large in the northern hemisphere (15 events) than in the southern hemisphere (7 events),
as displayed by Figure \ref{loc}a.
This implies that the GCR density minimum region propagating radially outward from the CME eruption on the sun passed north of Earth in most of the $E$-events,
canceling out with the north-south asymmetry of the density depression mentioned above.
The IMF sector polarity during FDs may also affect the $G_z$ distribution as mentioned in subsection \ref{sec:ev2},
but we have confirmed that the IMF sector dependence has only a minor effect on the average $G_z$ distribution in Figure \ref{ssGew}f,
by performing the correction for the IMF sector dependence described in the next section.

\subsubsection{GCR density gradient perpendicular to the ecliptic plane}\label{sec:ssNS}
The latitudinal (north-south) distribution of GCR density behind IP-shocks has rarely been investigated.
This is partly because solar eruptions are limited in low- and mid-latitude regions on the sun (see Figures \ref{loc}a and \ref{loc}c),
prohibiting reliable analyses of the latitudinal distribution from the GCR density observed at Earth's orbit.
The three-dimensional gradient vector analyzed in this paper allows us
to deduce the latitudinal density distribution as well as the distribution in the ecliptic plane.
The north-south component of the density gradient, $G_z$, is expected to be southward or negative (northward or positive) in the $N$-events ($S$-events),
if the density minimum region passes north (south) of Earth
while propagating radially outward from the CME eruption in the northern (southern) hemisphere of the sun.\par
It is noted, however, that the sector polarity of the IMF ({\it away} or {\it toward}) also has to be taken into account when we analyze $G_z$,
because the drift model of the large-scale GCR transport in the heliosphere predicts a persistent latitudinal gradient
which depends on the IMF sector polarity.
The drift model \citep{Kot82,Kot83} predicts a spatial distribution of the GCR density having a local maximum close to the heliospheric current sheet (HCS) \citep{Wil65}
in the ``negative'' polarity period of the solar polar magnetic field (also referred as the $A<0$ epoch) when the IMF directs toward (away from) the sun above (below) the HCS.
All SSC events before 2012 in Table \ref{tb:ev} are recorded in an $A<0$ epoch.
The density distribution in an $A>0$ epoch (period from 2013 in Table \ref{tb:ev})
when the IMF directs away from (toward) the sun above (below) the HCS, on the other hand, is predicted to have a minimum close to the HCS.
The drift model thus predicts positive (negative) $G_z$ in {\it away} ({\it toward}) IMF sectors
regardless of $A>0$ or $A<0$ epoch.
This drift model prediction of $G_z$ has been actually confirmed by previous analyses of the GMDN and NM data \citep{Che93,Oka08,Fus10a,Mun14,Koz14}.\par
\begin{figure}
   \plotone{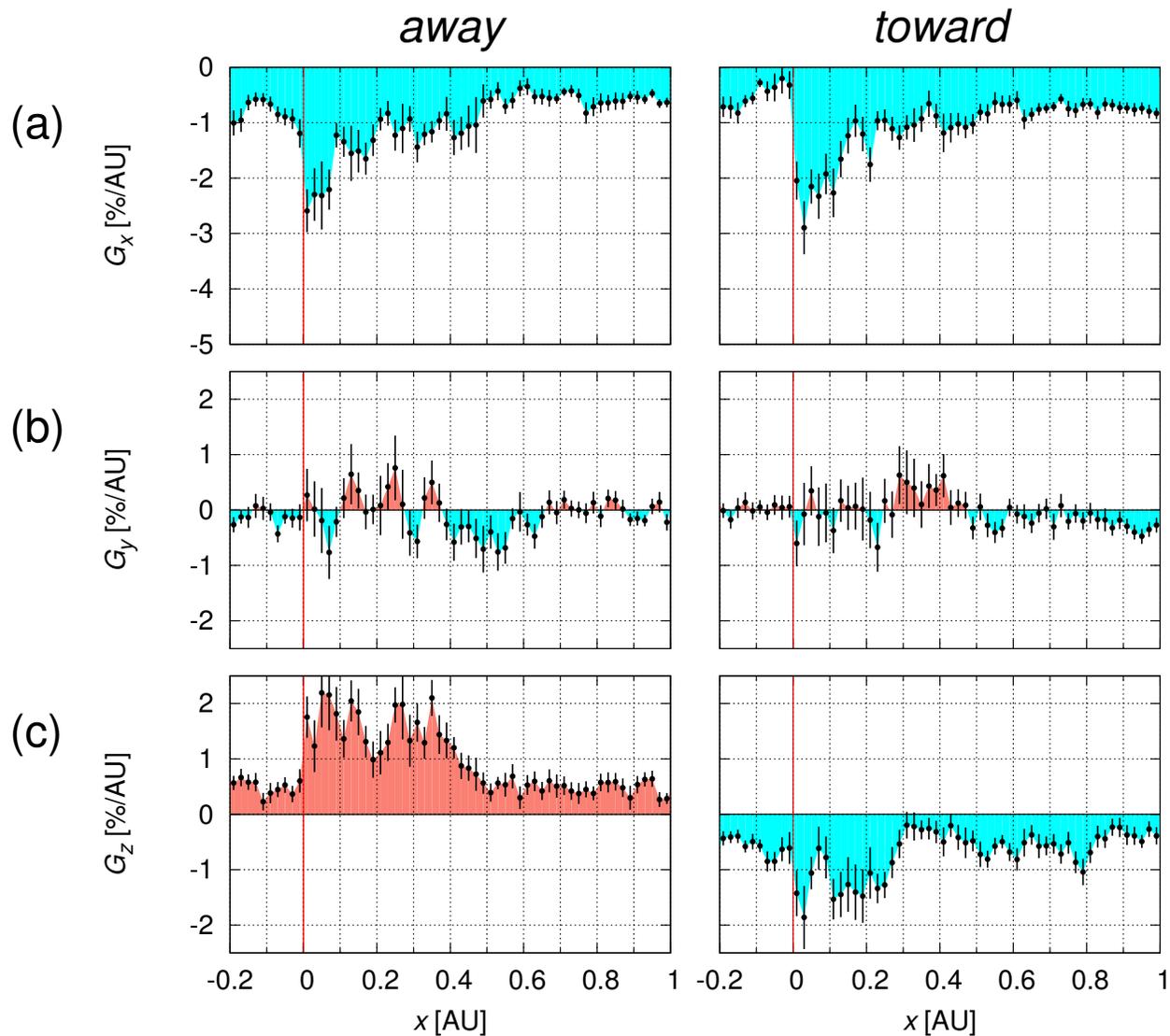}
   \caption{Averages of the superposed spatial distributions of (a) $G_x$, (b) $G_y$ and (c) $G_z$ in the same manner as Figure \ref{ssGew}.
   The left panels display the average distributions in the {\it away} IMF sector while the right panels are in the {\it toward} IMF sector.
   For the superposition in this figure, we used only the central events, as well as Figures \ref{ssI} and \ref{ssGew}.}
   \label{ssSec}
\end{figure}

Figure \ref{ssSec} shows the superposed density gradient of 45 central events in {\it away} and {\it toward} IMF sectors.
In producing this figure, IMF sector polarity is designated referring to
the hourly mean magnetic field ${\bm B}(t)$ in ACE or WIND data as described in subsection \ref{sec:ev1}.
The sector polarity is defined on an hourly basis in each event,
so hourly density gradients in an event are separated into two IMF sectors,
in cases where a sector boundary crossing is recorded during the event.
It is clear in Figure \ref{ssSec}c that the average $G_z$ is positive in the {\it away} sector (left panel)
while it is negative in the {\it toward} sector (right panel),
in accord with the drift model prediction described above.
The average distributions of $G_x$ and $G_y$ in Figures \ref{ssSec}a and  \ref{ssSec}b
do not show such a significant dependence on the IMF sector polarity.
It is also seen that the magnitude of $G_z$ is enhanced behind IP-shock ($x>0$),
i.e. the positive (negative) $G_z$ in the {\it away} ({\it toward}) sector is enhanced up to 3-5 times of that ahead the shock ($x<0$).\par
\begin{figure}
   \epsscale{0.9}
   \plotone{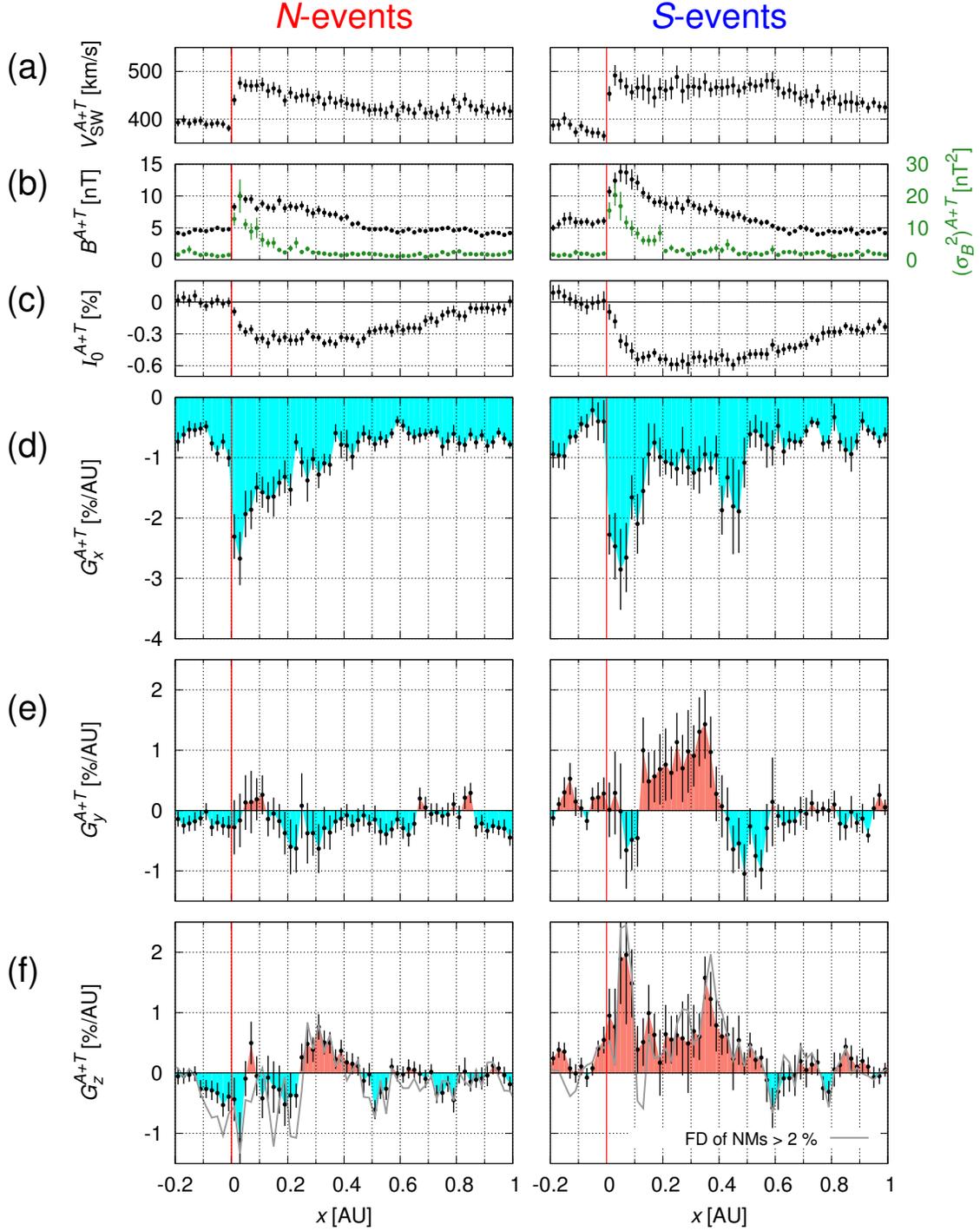}
   \caption{\footnotesize{Averages of the superposed spatial distributions in the (left) $N$-events and (right) $S$-events:
      (a-c) solar wind speed ($V_{\rm SW}^{A+T}$), magnetic field magnitude ($B^{A+T}$) and variance $\left((\sigma_B^2)^{A+T}\right)$ and GCR density ($I_0^{A+T}$)
      in the same manner as Figures \ref{ssI}a, \ref{ssI}b and \ref{ssI}d,
      and (d-f) three GSE components of the GCR density gradient ($G_x^{A+T}$, $G_y^{A+T}$, $G_z^{A+T}$).
      Each distributions in this figure is corrected for the IMF sector polarity dependence in Figure \ref{ssSec} by equations (\ref{eq:N}) and (\ref{eq:S}) (see text).
      The format is the same as Figure \ref{ssI}.}}
   \label{ssGns}
\end{figure}

In order to correct $G_z$ in the $N$- and $S$-events for the sector dependence mentioned above,
we first calculate the average $G_z$ in each IMF sector, respectively for the $N$- and $S$-events.
We then calculate the average $G_z$s in the $N$- and $S$-events, as
\begin{eqnarray}
   G_z^{A+T}(N) &=& \frac{G_z^A(N) + G_z^T(N)}{2} \label{eq:N}\\
   G_z^{A+T}(S) &=& \frac{G_z^A(S) + G_z^T(S)}{2} \label{eq:S}
\end{eqnarray}
where $G_z^A(N)$ and $G_z^T(N)$ ($G_z^A(S)$ and $G_z^T(S)$) are average $G_z$s
in the {\it away} and {\it toward} sectors in the $N$-events ($S$-events), respectively.
We present spatial distributions of the derived $G_z^A(N)$, $G_z^T(N)$, $G_z^A(S)$, and $G_z^T(S)$ in Section \ref{sec:GzSec}.
Figures \ref{ssGns}a-\ref{ssGns}e show the distributions of the solar wind speed ($V_{\rm SW}^{A+T}$),
IMF magnitude ($B^{A+T}$) and variance ($(\sigma_B^2)^{A+T}$), GCR density ($I_0^{A+T}$), and ecliptic components ($G_x^{A+T}, G_y^{A+T}$) of the density gradient,
all corrected for the IMF sector dependence in the same manner as equations (\ref{eq:N}) and (\ref{eq:S}).
Black points in the left (right) panel of Figure \ref{ssGns}f display the $G_z^{A+T}(N)$ ($G_z^{A+T}(S)$) distribution with
errors calculated from standard errors of $G_z^A(N)$ and $G_z^T(N)$ ($G_z^A(S)$ and $G_z^T(S)$) in equation (\ref{eq:N}) (equation (\ref{eq:S})).\par
It is clear in the right panel of Figure \ref{ssGns}f that the positive (northward) gradient is enhanced in the $S$-events.
This is qualitatively consistent with a density minimum region propagating radially outward from the CME eruption on the sun.
A negative $G_z$ enhancement in the $N$-events shown by black points in the left panel of Figure \ref{ssGns}f
is unclear compared with the positive enhancement in the $S$-events.
Durations or magnitudes of the enhancements in the solar wind speed (Figure \ref{ssGns}a),
IMF magnitude (Figure \ref{ssGns}b), and GCR density depression (Figure \ref{ssGns}c) are clearly shorter or smaller in the $N$-events than in the $S$-events,
indicating that the $N$-events were weaker than the $S$-events.
This may result in less significant negative enhancement of $G_z$ in the $N$-events when compared with the positive enhancement in the $S$-events.
In the intense events in which the maximum density depression in FDs derived from NM data exceed 2 \% (gray curve in Figure \ref{ssGns}f),
we can see that in $N$-events the negative $G_z$ enhancement behind the IP-shocks in 0 AU $<x<$ +0.2 AU is larger than the black points.
We note that $G_z$ in Figure \ref{ssGns}f shows simultaneous enhancements
in 0 AU $<x<$ +0.1 AU with opposite signs in $N$- and $S$-events,
which may possibly be related to the sheath structure between the IP-shocks
and the CME ejecta as well as $G_y$ in Figure \ref{ssGew}e,
although this is unclear due to the big error bars.\par
The GSE-$y$ component of the density gradient, $G_y$ in Figure \ref{ssGns}e, shows a positive enhancement in the $S$-events,
while the $N$-events are dominated by a negative $G_y$.
This can be attributed to the east-west asymmetry of the $N$- and $S$-event numbers.
In the central region of the southern hemisphere on the sun, 12 CMEs erupted from the western ($0^\circ \le \phi \le +45^\circ$) region
while 7 CMEs erupted from the eastern ($-45^\circ \le \phi < 0^\circ$) region, as seen in the event number in Figure \ref{loc}a.
This indicates that the CME eruptions associated with the $S$-events are dominated by those on the western side on the sun,
which may cause the density minimum regions passing west of Earth and the positive $G_y$ enhancement in the right panel of Figure \ref{ssGns}e.
CME eruptions from the northern hemisphere on the sun, on the other hand,
have a larger event number (15 events) in the eastern region than in the western region (11 events),
possibly resulting in the negative $G_y$ in the left panel of Figure \ref{ssGns}e.

\section{Summary and conclusion}\label{sec:Sum}
Most of the previous studies of FDs analyze the temporal variation of a single detector count rate
as monitoring the GCR density, or the isotropic intensity at Earth.
Cosmic ray intensity observed with a ground-based detector, however, includes contributions not only from the density, but also from the GCR anisotropy simultaneously.
Only a worldwide detector network, such as the GMDN, allows us to observe the cosmic ray density and anisotropy separately with a sufficient time resolution.\par
It has been shown in a series of papers that the GCR density gradient deduced from the anisotropy observed with the GMDN
is useful to infer the three dimensional geometry of the cylinder-type depleted region in the MFR \citep{Mun03,Mun06,Kuw04,Kuw09,Roc14}.
In this paper, we present a superposition analysis of dozens of FDs in Table \ref{tb:ev} observed since 2006 when the full-scale GMDN started operation.
We particularly analyze the GCR density gradient deduced from the anisotropy together with the density in FDs recorded following the IP-shocks,
each caused by an identified solar eruption.
By analyzing the superposed density and gradient in FDs caused by eastern, western, northern and southern eruptions on the sun,
i.e. the $E$-, $W$-, $N$-, and $S$-events respectively, we deduced the average spatial distribution of GCRs in FDs.\par
We found two distinct modulations of GCR density in FDs.
One is in the magnetic sheath region which extends over $\sim 0.1$ AU in GSE-$x$ behind IP-shocks.
The density depression in the sheath region is larger in the western flank of IP-shocks than in the eastern flank,
because the stronger compressed IMF in the western flank shields more GCRs from outside as suggested by \citet{Hau65}.\par
The other modulation is in the central CME region behind the sheath and causes the additional density depression in $x>+0.1$ AU.
This is attributed to an adiabatic expansion of the MFR formed in the central region of the CME.
The density minimum at the longitudinal center behind the IP-shock,
which is caused by the CME ejecta or MFR, was confirmed from the negative and positive enhancements of $G_y$ in the $E$- and $W$-events, respectively.
The negative and positive $G_z$ enhancements in the $N$- and $S$-events,
indicating the density minimum at the latitudinal center behind IP-shock,
are also seen when $G_z$ is corrected for the asymmetry in the {\it away} and {\it toward} IMF sectors
(that is, above and below the HCS) predicted by the drift model.
We also note that the centered density minimum was seen not only in the central CME region but also in the sheath region.\par
By comparing the density depressions observed with the GMDN and NMs,
we confirmed that the rigidity spectrum of the density depression
is consistent overall with a soft power-law spectrum $P^\Gamma$ with $\Gamma \sim -0.8$ as seen in Figure \ref{ssI}f.
It was also found that the modulation in the central CME region has
a softer rigidity spectrum than the modulation in the magnetic sheath.
This may be related to a difference between GCR diffusion coefficients in the ordered magnetic field of the MFR and
in the turbulent IMF in the sheath region.
The rigidity spectrum getting softer during the recovery phase in $E$-events
implies that the density depression recoveries faster in $\sim 60$ GV GCRs than in $\sim 10$ GV GCRs,
while such a recovery is not seen in the $W$-events due to the longer duration of the solar wind speed enhancement.
Previous studies \citep{Bie98,Mun03,Mun06,Kuw04,Kuw09,Roc14} analyzed the GMDN and NM data separately,
but the combined analyses of these data sets, as presented in the present paper,
can provide us with important information on the rigidity dependence of GCR modulation in space weather.
We plan to make such analyses in the near future by using the data observed
with the world network of NMs and the GMDN.
We also note a recent interesting paper by \citet{Ruf16}
reporting the rigidity dependence derived from a single NM observation.\par
In addition to the asymmetry in the {\it away} and {\it toward} IMF sectors,
$G_z$ also shows negative and positive enhancements behind IP-shocks as shown in Figure \ref{ssSec}.
An enhanced longitudinal component of IMF in the sheath behind IP-shocks is expected to cause a latitudinal $\nabla B$ drift \citep{Sar89}
and possibly enhance the latitudinal density gradient
which changes sign in {\it away} and {\it toward} IMF sectors as the observed $G_z$.\par
The average spatial distribution of the GCR density in FD presented in this paper demonstrates that
the observations of high energy GCR density and anisotropy with the GMDN and NMs provide us with good tools
also for studying the space weather in the region of IP-shocks.

\section*{Acknowledgement}
This work is supported in part by the joint research programs of the Institute for Space-Earth Environmental Research (ISEE), Nagoya University and
the Institute for Cosmic Ray Research (ICRR), University of Tokyo.
The observations are supported by Nagoya University with the Nagoya muon detector,
by CNPq, CAPES, INPE, and UFSM with the S\~ao Martinho da Serra muon detector, and
by the Australian Antarctic Division with the Hobart muon detector.
Observations by the S\~ao Martinho da Serra muon detector are also supported by
the S\~ao Paulo Research Foundation (FAPESP) grants
\#2014/24711-6, \#2013/03530-0, \#2013/02712-8, \#2012/05436-9, \#2012/20594-0, and \#2008/08840-0.
The Kuwait muon detector is supported by project SP01/09 of the Research Administration of Kuwait University.
Neutron monitors of the Bartol Research Institute are supported by the National Science Foundation grant ATM-0000315.
The SSC list was obtained from Helmholtz-Centre Potsdam - GFZ via the website at http://www.gfz-potsdam.de.
The SWPC events lists were obtained via the website at http://www.swpc.noaa.gov.
We thank the ACE SWEPAM instrument team and the ACE Science Center for providing the ACE data.
The WIND spacecraft data were obtained via the NASA homepage.
The SW news was provided by National Institute of Information and Communications Technology (NICT) until 2009 and is currently provided by NIT, Kagoshima College.
We thank M. Shinohara for updating the SW news every day.
We also thank to Dr. V. Yanke of the Institute of Terrestrial Magnetism,
   Ionosphere and Radio Wave Propagation RAS (IZMIRAN)
for providing us with the GMDN data in 2009 corrected for the atmospheric temperature effect.

\appendix
\section{GCR density gradient inferred from the density distribution}\label{sec:Gx}
For the first time, we discuss a structure of the GCR depleted region behind IP-shocks using the density gradient derived from the first order anisotropy.
It is thus important to confirm the consistency between the gradient and the density which has been analyzed by most of the earlier works.
We infer the GSE-$x$ component of the density gradient, $G_x$, not from the anisotropy but from the observed density in this section for the comparison.\par
\begin{figure}
   \epsscale{1}
   \plotone{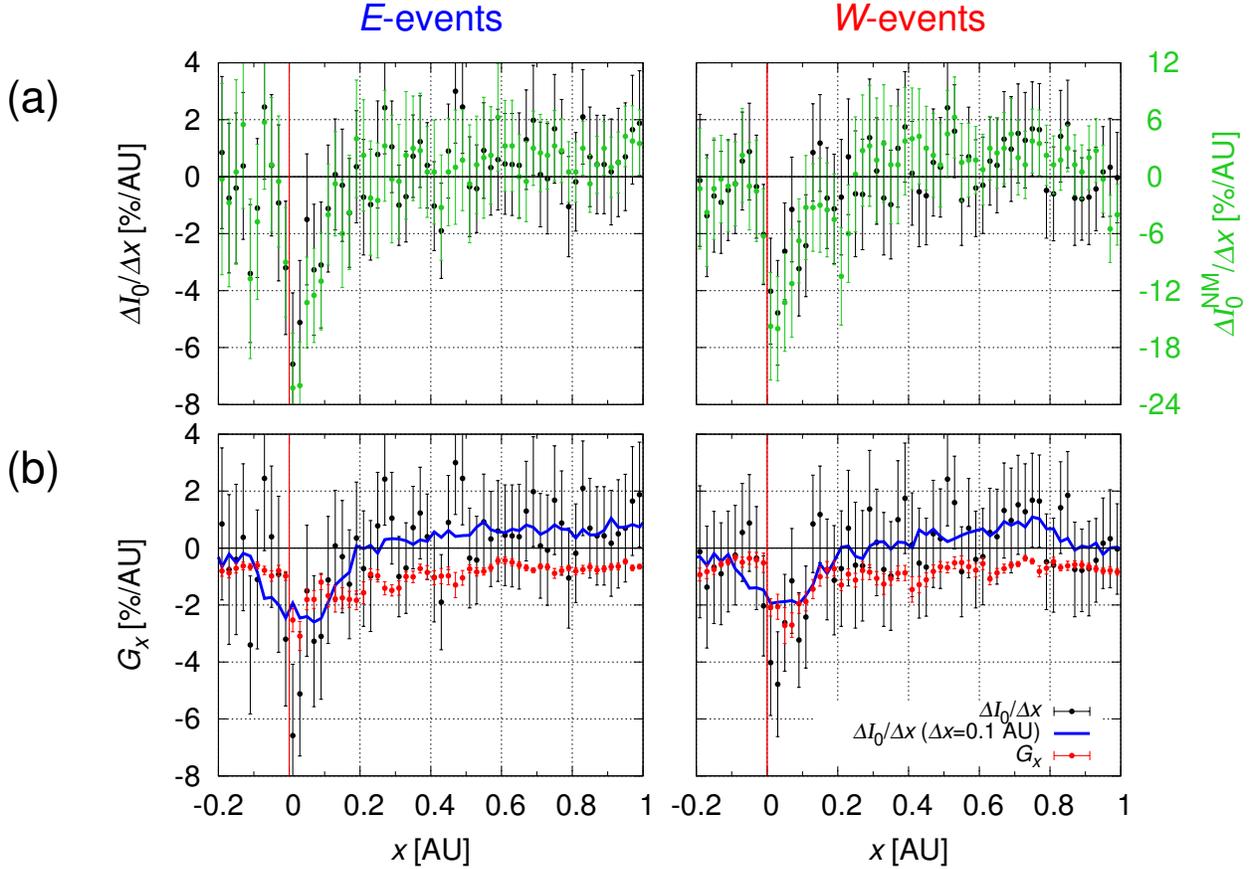}
   \caption{GSE-$x$ components of the density gradient in (left) $E$- and (right) $W$- events inferred
      from the density distributions observed with the GMDN.
   Black points in panels (a) and (b) represent the gradient calculated with $\Delta x = 0.02$ AU,
   while a blue curve in panel (b) is the gradient with $\Delta x = 0.1$ AU (see text).
   A green point in panel (a) is the density gradient inferred from the density distributions
   observed with NMs, calculated with $\Delta x = 0.02$ AU.
   Red points in panel (b) is the same as the black points in Figure \ref{ssGew}d,
an average of the superposed gradient derived from the anisotropy.}
   \label{ssGx}
\end{figure}

We calculate the density gradient $\Delta I_0(x)/\Delta x$ from the superposed $I_0(x)$ shown by black points in Figure \ref{ssI}d, as
\begin{equation}
   \frac{\Delta I_0(x)}{\Delta x} = \frac{I_0(x + \Delta x) - I_0(x - \Delta x)}{2\Delta x}
   \label{eq:I2G}
\end{equation}
where $\Delta x$ is set at 0.02 AU as an ad hoc choice.
Black points in Figures \ref{ssGx}a and  \ref{ssGx}b display $\Delta I_0(x)/\Delta x$.
A green point in Figure \ref{ssGx}a shows $\Delta I_0^{\rm NM}(x)/\Delta x$,
the density gradient inferred from the density distribution ($I_0^{\rm NM}(x)$) observed with NMs,
which is displayed by black points in Figure \ref{ssI}e.
It is seen that $\Delta I_0(x)/\Delta x$ and $\Delta I_0^{\rm NM}(x)/\Delta x$
are in good agreement with a similar GSE-$x$ range (0 AU $<x<$ +0.2 AU) of their negative enhancement,
while the magnitude of the enhancement is three times larger in $\Delta I_0^{\rm NM}(x)/\Delta x$ than in $\Delta I_0(x)/\Delta x$.\par
The red points in Figure \ref{ssGx}b are the GSE-$x$ component ($G_x$) of the density gradient derived from the anisotropy,
the same as the black points in Figure \ref{ssGew}d.
We cannot confirm a consistency between $\Delta I_0(x)/\Delta x$ and $G_x$ in Figure \ref{ssGx}b due to the large fluctuation of $\Delta I_0(x)/\Delta x$,
but a negative enhancement of $\Delta I_0(x)/\Delta x$ in 0 AU $<x<$ +0.2 AU seems roughly comparable with $G_x$.\par
It is noted that the density gradient, or the anisotropy,
can be regarded as reflecting a global spatial structure over $\sim 0.1$ AU
because $\sim 60$ GV GCRs have a Larmor radius of $\sim 0.2$ AU in the IMF of $B=7$ nT.
We change, therefore, the spatial interval $\Delta x$ in equation (\ref{eq:I2G}) to $\Delta x = 0.1$ AU.
The blue curve in Figure \ref{ssGx}b represents $\Delta I_0(x)/\Delta x$ derived from equation (\ref{eq:I2G}) with $\Delta x = 0.1$ AU.
The magnitude of the negative enhancement in the blue curve is fairly consistent with $G_x$ (red points),
implying that the density gradient derived from the anisotropy
reflects a global structure over a spatial scale comparable to the Larmor radius.
We also see some differences between $\Delta I_0(x)/\Delta x$ and $G_x$,
e.g. $\Delta I_0(x)/\Delta x$ (blue curve) shows a negative enhancement
starting before the SSC onset ($x<0$ AU), but this is obviously an artificial
effect of the central derivative with a large $\Delta x$ in equation (\ref{eq:I2G}).
The $G_x$ deduced from the GMDN (red points), on the other hand,
shows the enhancement immediately after the SSC onset.

\section{North-south component of the density gradient in each IMF sector}\label{sec:GzSec}
\begin{figure}
   \epsscale{1}
   \plotone{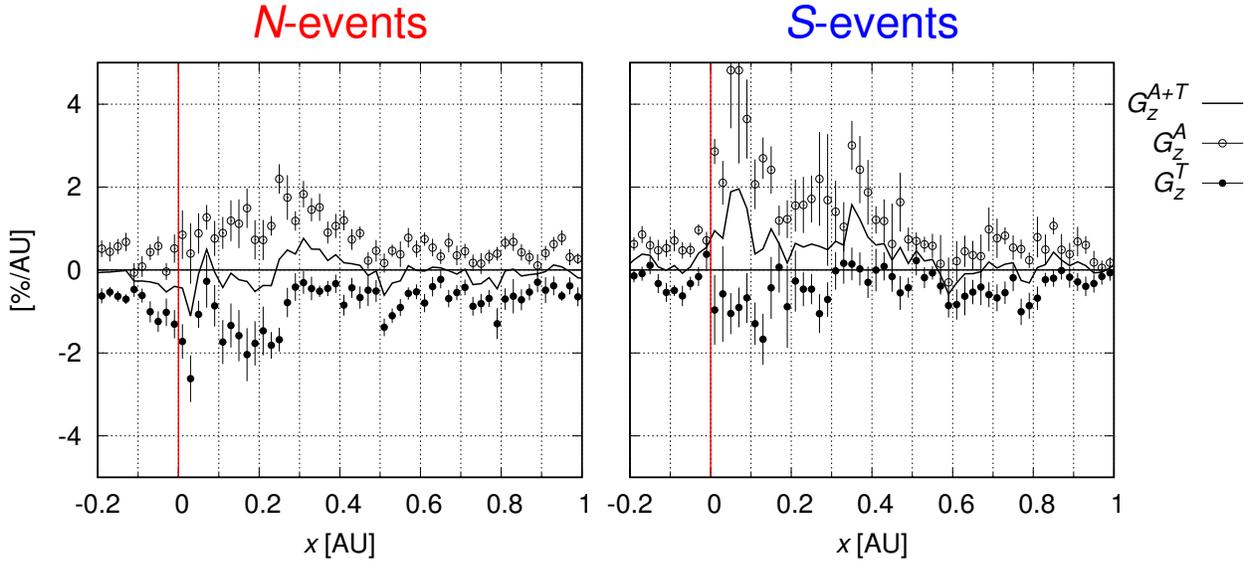}
   \caption{Average spatial distributions of the north-south components ($G_z^A$ and $G_z^T$) of density gradient in {\it away} and {\it toward} IMF sectors.
   Open and solid circles in the left (right) panel display $G_z^A(N)$ and $G_z^T(N)$ ($G_z^A(S)$ and $G_z^T(S)$) in $N$-events ($S$-events), respectively.
   A black line in the left (right) panel represents an average $G_z^{A+T}(N)$ ($G_z^{A+T}(S)$) of $G_z^A(N)$ and $G_z^T(N)$ ($G_z^A(S)$ and $G_z^T(S)$),
   i.e. the same as black points in the left (right) panel of Figure \ref{ssGns}f.
   For definitions of $G_z^A(N)$, $G_z^T(N)$, and $G_z^{A+T}(N)$ ($G_z^A(S)$, $G_z^T(S)$, and $G_z^{A+T}(S)$), see Section \ref{sec:ssNS}.}
   \label{ssGnsSec}
\end{figure}

The left (right) panel of Figure \ref{ssGnsSec} displays average spatial distributions of
the north-south components of density gradient ($G_z$)
in {\it away} and {\it toward} IMF sectors in $N$-events ($S$-events),
i.e. $G_z^A(N)$ and $G_z^T(N)$ ($G_z^A(S)$ and $G_z^T(S)$) in equation (\ref{eq:N}) (equation (\ref{eq:S})),
by open and solid circles, respectively.
We can see that the positive (negative) $G_z$ in {\it away} ({\it toward}) sector is enhanced behind IP-shock
in both of the $N$- and $S$-events as discussed in Section \ref{sec:ssNS}.
It is also noted that the positive enhancement of $G_z^A$ is larger than
the negative enhancement of $G_z^T$ in $S$-events,
resulting in the positive enhancement of $G_z^{A+T}$ in $S$-events.
This implies that a positive $G_z$, which arises from the density minimum region propagating
radially outward from the southern region on the sun,
is superposed on the both of the positive and negative $G_z$s in {\it away} and {\it toward} IMF sectors.
In $N$-events, we can also see that
the negative enhancement of $G_z^T$ is slightly larger than the positive enhancement of $G_z^A$,
while this is unclear compared with $S$-events.
This results in the small negative enhancement of $G_z^{A+T}$ in $N$-events, as discussed in Section \ref{sec:ssNS}.

\clearpage
\begin{deluxetable}{rcccccccccccccccc}
   \tabletypesize{\scriptsize}
   \rotate
   \tablewidth{0pt}
   \tablecaption{List of SSC events associated with solar eruptions}
   \tablehead{
      & \multicolumn{2}{c}{SSC onset} &
      & \multicolumn{3}{c}{FD (GMDN)$^a$} &
      & \multicolumn{3}{c}{FD (NMs)$^b$} &
      & \multicolumn{5}{c}{Associated event on the sun}\\\cline{2-3} \cline{5-7} \cline{9-11} \cline{13-17}
      & &  &
      & & & \colhead{max.} &
      & & & \colhead{max.} &
      & & & & \colhead{X-ray} & \colhead{heliographic}\\
      \colhead{No.} & \colhead{date} & \colhead{time} &
      & \colhead{date} & \colhead{time} & \colhead{[\%]} &
      & \colhead{date} & \colhead{time} & \colhead{[\%]} &
      & \colhead{type$^c$} & \colhead{date$^d$} & \colhead{time$^d$} & \colhead{class$^d$} & \colhead{lat. \& long.$^d$}}
   \startdata
          $^{\dagger}$1 & 2006/01/01 & 14:05 & &         -- &    -- &   -- & & 2006/01/01 & 19:30 & 0.2 & & FLA & 2005/12/29 & 21:11 & C1.1 & N11E17\\
                      2 & 2006/07/09 & 21:36 & & 2006/07/12 & 04:30 & 1.29 & & 2006/07/11 & 08:30 & 4.0 & & FLA & 2006/07/06 & 08:23 & M2.5 & S11W32\\
                      3 & 2006/08/19 & 11:30 & & 2006/08/20 & 10:30 & 0.59 & & 2006/08/20 & 10:30 & 2.9 & & FLA & 2006/08/16 & 16:17 & C3.6 & S14W13$^{\ast}$\\
                      4 & 2006/12/08 & 04:35 & & 2006/12/08 & 21:30 & 0.38 & & 2006/12/11 & 03:30 & 3.4 & & FLA & 2006/12/05 & 10:38 & X9.0 & S07E79\\
                      5 & 2006/12/14 & 14:14 & & 2006/12/15 & 02:30 & 2.34 & & 2006/12/15 & 00:30 & 7.4 & & FLA & 2006/12/13 & 02:34 & X3.4 & S06W24\\
                      6 & 2006/12/16 & 17:55 & & 2006/12/16 & 23:30 & 0.06 & & 2006/12/16 & 23:30 & 0.4 & & FLA & 2006/12/14 & 22:17 & X1.5 & S06W46\\
                      7 & 2007/05/21 & 23:20 & & 2007/05/22 & 11:30 & 0.51 & & 2007/05/22 & 14:30 & 1.6 & & DSF & 2007/05/19 & 12:31 & B9.5 & N07W06\\
         $^{\ddagger}$8 & 2007/11/19 & 18:11 & & 2007/11/20 & 01:30 & 0.31 & & 2007/11/20 & 04:30 & 2.5 & &  -- &         -- &    -- &   -- &     --\\
                      9 & 2008/04/30 & 15:57 & & 2008/04/30 & 21:30 & 0.11 & & 2008/05/01 & 19:30 & 0.9 & & FLA & 2008/04/26 & 14:00 & B3.8 & N08E09\\
        $^{\ddagger}$10 & 2008/12/16 & 11:59 & & 2008/12/17 & 13:30 & 0.45 & & 2008/12/17 & 13:30 & 2.9 & & DSF &         -- &    -- &   -- &     --\\
 $^{\dagger\ddagger}$11 & 2009/10/22 & 00:17 & &         -- &    -- &   -- & & 2009/10/23 & 01:30 & 1.4 & &  -- &         -- &    -- &   -- &     --\\
                     12 & 2010/04/05 & 08:26 & & 2010/04/06 & 13:30 & 0.75 & & 2010/04/06 & 02:30 & 2.6 & & FLA & 2010/04/03 & 00:24 & B7.4 & S25W00\\
                     13 & 2010/04/11 & 13:04 & & 2010/04/12 & 03:30 & 0.32 & & 2010/04/12 & 07:30 & 1.4 & & FLA & 2010/04/08 & 03:25 & B3.7 & N24E12$^{\ast}$\\
        $^{\ddagger}$14 & 2010/05/28 & 02:58 & & 2010/05/29 & 08:30 & 0.47 & & 2010/05/30 & 04:30 & 1.9 & & FLA & 2010/05/24 & 14:46 & B1.1 &     --\\
                     15 & 2010/08/03 & 17:40 & & 2010/08/04 & 10:30 & 0.76 & & 2010/08/04 & 05:30 & 3.4 & & FLA & 2010/08/01 & 08:26 & C3.2 & N20E36\\
                     16 & 2010/12/19 & 21:32 & & 2010/12/20 & 02:30 & 0.18 & & 2010/12/22 & 10:30 & 0.7 & & DSF & 2010/12/16 & 04:27 & B7.4 & N18E27$^{\ast}$\\
                     17 & 2011/02/18 & 01:30 & & 2011/02/18 & 11:30 & 1.37 & & 2011/02/18 & 12:30 & 4.6 & & FLA & 2011/02/15 & 01:56 & X2.2 & S21W15$^{\ast}$\\
                     18 & 2011/03/10 & 06:32 & & 2011/03/12 & 14:30 & 0.73 & & 2011/03/11 & 08:30 & 2.7 & & FLA & 2011/03/07 & 20:12 & M3.7 & N24W58$^{\ast}$\\
                     19 & 2011/04/06 & 09:33 & & 2011/04/08 & 22:30 & 1.23 & & 2011/04/08 & 04:30 & 1.6 & & FLA & 2011/04/03 & 05:19 & C1.2 & N15W15$^{\ast}$\\
                     20 & 2011/04/18 & 06:52 & & 2011/04/18 & 07:30 & 0.03 & & 2011/04/19 & 09:30 & 0.6 & & FLA & 2011/04/15 & 17:11 & M1.3 & N14W19\\
                     21 & 2011/06/04 & 20:44 & & 2011/06/05 & 09:30 & 1.04 & & 2011/06/05 & 05:30 & 3.2 & & FLA & 2011/06/01 & 16:56 & C4.1 & S20E20\\
                     22 & 2011/06/10 & 08:47 & & 2011/06/10 & 23:30 & 0.97 & & 2011/06/10 & 18:30 & 1.8 & & FLA & 2011/06/07 & 06:29 & M2.5 & S21W54\\
                     23 & 2011/06/17 & 02:39 & & 2011/06/17 & 13:30 & 1.18 & & 2011/06/18 & 02:30 & 2.9 & & DSF & 2011/06/14 & 21:42 & M1.3 & N15E77\\
                     24 & 2011/07/11 & 08:51 & & 2011/07/11 & 16:30 & 1.20 & & 2011/07/12 & 02:30 & 3.1 & & FLA & 2011/07/09 & 00:48 &   -- & S18E11$^{\ast}$\\
                     25 & 2011/08/05 & 17:51 & & 2011/08/06 & 22:30 & 1.07 & & 2011/08/06 & 13:30 & 4.4 & & FLA & 2011/08/03 & 13:50 & M6.0 & N16W30\\
                     26 & 2011/09/17 & 03:43 & & 2011/09/18 & 13:30 & 0.46 & & 2011/09/18 & 10:30 & 2.3 & & FLA & 2011/09/13 & 23:33 &   -- & N23W03\\
                     27 & 2011/09/25 & 11:45 & & 2011/09/26 & 01:30 & 0.41 & & 2011/09/26 & 02:30 & 1.4 & & FLA & 2011/09/22 & 10:57 & X1.4 & N13E78\\
                     28 & 2011/09/26 & 12:35 & & 2011/09/28 & 11:30 & 0.99 & & 2011/09/27 & 06:30 & 4.5 & & FLA & 2011/09/24 & 13:20 & M7.1 & N13E51\\
                     29 & 2011/10/05 & 07:36 & & 2011/10/05 & 18:30 & 0.85 & & 2011/10/05 & 21:30 & 2.4 & & FLA & 2011/10/02 & 00:49 & M3.9 & N09W12\\
                     30 & 2011/10/24 & 18:31 & & 2011/10/25 & 09:30 & 1.17 & & 2011/10/25 & 06:30 & 5.9 & & FLA & 2011/10/22 & 10:18 & M1.3 & N25W77\\
                     31 & 2011/11/12 & 05:59 & & 2011/11/13 & 04:30 & 0.40 & & 2011/11/13 & 15:30 & 1.7 & & FLA & 2011/11/09 & 13:35 & M1.1 & N18E26$^{\ast}$\\
                     32 & 2011/11/28 & 21:50 & & 2011/11/29 & 14:30 & 0.40 & & 2011/11/30 & 10:30 & 2.4 & & FLA & 2011/11/26 & 07:10 & C1.2 & N08W39$^{\ast}$\\
                     33 & 2011/12/18 & 19:03 & & 2011/12/22 & 11:30 & 0.63 & & 2011/12/22 & 13:30 & 1.8 & & FLA & 2011/12/13 & 23:34 & C4.8 & S19W84\\
                     34 & 2012/01/22 & 06:12 & & 2012/01/22 & 23:30 & 0.76 & & 2012/01/23 & 10:30 & 3.4 & & FLA & 2012/01/16 & 04:44 & C6.5 & N27E61$^{\ast}$\\
                     35 & 2012/01/24 & 15:03 & & 2012/01/24 & 16:30 & 0.75 & & 2012/01/24 & 17:30 & 3.3 & & FLA & 2012/01/23 & 04:04 & M8.7 & N28W21\\
                     36 & 2012/01/30 & 16:24 & & 2012/02/01 & 12:30 & 0.69 & & 2012/02/01 & 07:30 & 3.4 & & FLA & 2012/01/27 & 18:51 & X1.7 & N27W71\\
                     37 & 2012/02/26 & 21:40 & & 2012/02/29 & 12:30 & 1.90 & & 2012/02/28 & 14:30 & 3.3 & & DSF & 2012/02/24 & 02:25 &   -- & N32E38\\
                     38 & 2012/03/07 & 04:20 & & 2012/03/08 & 10:30 & 0.82 & & 2012/03/08 & 10:30 & 2.8 & & FLA & 2012/03/05 & 03:48 & X1.1 & N17E52\\
                     39 & 2012/03/08 & 11:03 & & 2012/03/09 & 00:30 & 1.98 & & 2012/03/09 & 07:30 & 9.8 & & FLA & 2012/03/07 & 00:17 & X5.4 & N17E27\\
                     40 & 2012/03/12 & 09:15 & & 2012/03/13 & 01:30 & 1.03 & & 2012/03/13 & 04:30 & 4.8 & & FLA & 2012/03/09 & 03:53 & M6.3 & N17W01$^{\ast}$\\
                     41 & 2012/03/15 & 13:07 & & 2012/03/15 & 23:30 & 0.28 & & 2012/03/15 & 18:30 & 1.1 & & FLA & 2012/03/13 & 17:41 & M7.9 & N18W61$^{\ast}$\\
        $^{\ddagger}$42 & 2012/04/23 & 03:20 & & 2012/04/26 & 11:30 & 1.27 & & 2012/04/26 & 05:30 & 3.4 & &  -- & 2012/04/19 & 15:15 & C1.8 &     --\\
                     43 & 2012/05/21 & 19:37 & & 2012/05/22 & 20:30 & 0.09 & & 2012/05/22 & 05:30 & 0.7 & & FLA & 2012/05/17 & 01:34 & M5.1 & N11W76\\
                     44 & 2012/06/16 & 20:20 & & 2012/06/17 & 05:30 & 1.15 & & 2012/06/17 & 04:30 & 4.5 & & FLA & 2012/06/13 & 13:41 & M1.2 & S16E18\\
                     45 & 2012/07/14 & 18:09 & & 2012/07/15 & 13:30 & 1.26 & & 2012/07/15 & 18:30 & 5.9 & & FLA & 2012/07/12 & 16:25 & X1.4 & S15W01\\
                     46 & 2012/08/02 & 10:50 & & 2012/08/02 & 15:30 & 0.50 & & 2012/08/03 & 08:30 & 0.3 & & FLA & 2012/07/28 & 20:58 & M6.1 & S25E54\\
                     47 & 2012/08/16 & 13:15 & & 2012/08/16 & 14:30 & 0.02 & & 2012/08/17 & 06:30 & 0.9 & & FLA & 2012/08/14 & 11:37 & C1.1 & N20W12$^{\ast}$\\
                     48 & 2012/09/03 & 12:13 & & 2012/09/05 & 10:30 & 1.48 & & 2012/09/05 & 10:30 & 4.6 & & FLA & 2012/08/31 & 20:43 & C8.4 & S19E42\\
                     49 & 2012/09/30 & 23:05 & & 2012/10/02 & 07:30 & 1.40 & & 2012/10/02 & 05:30 & 1.8 & & FLA & 2012/09/27 & 23:48 & C3.7 & N06W34\\
                     50 & 2012/10/08 & 05:16 & & 2012/10/10 & 03:30 & 0.20 & & 2012/10/10 & 14:30 & 1.9 & & FLA & 2012/10/05 & 07:30 & B7.8 & S22W30$^{\ast}$\\
        $^{\ddagger}$51 & 2012/10/31 & 15:39 & & 2012/11/01 & 15:30 & 0.25 & & 2012/11/01 & 07:30 & 1.6 & & DSF &         -- &    -- &   -- &     --\\
        $^{\ddagger}$52 & 2012/11/12 & 23:12 & & 2012/11/13 & 18:30 & 0.77 & & 2012/11/13 & 17:30 & 3.3 & & DSF & 2012/11/09 & 16:06 &   -- &     --\\
        $^{\ddagger}$53 & 2012/11/23 & 21:52 & & 2012/11/26 & 05:30 & 1.00 & & 2012/11/24 & 22:30 & 3.2 & & FLA & 2012/11/20 & 12:41 & M1.7 &     --\\
                     54 & 2013/02/16 & 12:09 & & 2013/02/18 & 13:30 & 1.37 & & 2013/02/18 & 06:30 & 2.4 & & DSF & 2013/02/13 & 03:00 &   -- & S28W54\\
                     55 & 2013/03/17 & 06:00 & & 2013/03/19 & 12:30 & 1.23 & & 2013/03/18 & 03:30 & 4.4 & & FLA & 2013/03/15 & 06:37 & M1.1 & N11E12\\
                     56 & 2013/04/13 & 22:55 & & 2013/04/16 & 11:30 & 1.13 & & 2013/04/15 & 13:30 & 3.4 & & FLA & 2013/04/11 & 07:10 & M6.5 & N09E12\\
                     57 & 2013/04/30 & 09:49 & & 2013/04/30 & 21:30 & 0.64 & & 2013/05/01 & 09:30 & 2.8 & & DSF & 2013/04/26 & 09:25 &   -- & S38W05\\
                     58 & 2013/05/18 & 01:10 & & 2013/05/18 & 19:30 & 0.57 & & 2013/05/18 & 22:30 & 1.7 & & FLA & 2013/05/15 & 01:40 & X1.2 & N12E64\\
                     59 & 2013/05/19 & 23:08 & & 2013/05/20 & 10:30 & 0.25 & & 2013/05/21 & 00:30 & 1.1 & & FLA & 2013/05/17 & 08:54 & M3.2 & N12E31\\
                     60 & 2013/05/24 & 18:10 & & 2013/05/26 & 04:30 & 0.66 & & 2013/05/26 & 06:30 & 3.0 & & FLA & 2013/05/22 & 13:22 & M5.0 & N15W70\\
        $^{\ddagger}$61 & 2013/05/31 & 16:18 & & 2013/06/04 & 00:30 & 0.49 & & 2013/06/01 & 01:30 & 1.4 & &  -- &         -- &    -- &   -- &     --\\
                     62 & 2013/06/27 & 14:38 & & 2013/06/28 & 08:30 & 0.76 & & 2013/06/28 & 05:30 & 2.5 & & FLA & 2013/06/23 & 20:53 & M2.9 & S15E62\\
        $^{\ddagger}$63 & 2013/10/02 & 01:55 & & 2013/10/02 & 07:30 & 0.88 & & 2013/10/02 & 07:30 & 3.1 & & DSF & 2013/09/29 & 23:39 & C1.2 &     --\\
        $^{\ddagger}$64 & 2013/12/13 & 13:22 & & 2013/12/14 & 11:30 & 0.38 & & 2013/12/14 & 13:30 & 1.0 & &  -- &         -- &    -- &   -- &     --\\
                     65 & 2014/02/07 & 17:05 & & 2014/02/07 & 21:30 & 0.14 & & 2014/02/08 & 16:30 & 0.7 & & FLA & 2014/02/04 & 03:58 & M5.2 & S14W06\\
                     66 & 2014/02/20 & 03:20 & & 2014/02/20 & 18:30 & 0.49 & & 2014/02/20 & 18:30 & 2.9 & & DSF & 2014/02/18 & 06:14 &   -- & S31E54\\
                     67 & 2014/02/27 & 16:50 & & 2014/02/28 & 23:30 & 0.94 & & 2014/02/28 & 17:30 & 3.9 & & FLA & 2014/02/25 & 00:47 & X4.9 & S12E82\\
                     68 & 2014/03/25 & 20:04 & & 2014/03/26 & 19:30 & 0.22 & & 2014/03/26 & 16:30 & 1.4 & & FLA & 2014/03/23 & 03:23 & C5.0 & S12E40\\
                     69 & 2014/04/20 & 10:56 & & 2014/04/20 & 19:30 & 0.58 & & 2014/04/20 & 23:30 & 1.2 & & FLA & 2014/04/18 & 13:03 & M7.3 & S15W36$^{\ast}$\\
                     70 & 2014/04/29 & 20:26 & & 2014/04/30 & 00:30 & 0.05 & & 2014/04/30 & 05:30 & 0.9 & & FLA & 2014/04/25 & 00:42 & X1.3 & S15W90\\
                     71 & 2014/06/23 & 23:08 & & 2014/06/24 & 11:30 & 0.05 & & 2014/06/24 & 15:30 & 0.3 & & DSF & 2014/06/19 & 09:15 &   -- & S01E15\\
                     72 & 2014/07/03 & 00:42 & & 2014/07/03 & 07:30 & 0.11 & & 2014/07/03 & 01:30 & 0.1 & & FLA & 2014/06/28 & 08:58 & C1.1 & N09E05\\
                     73 & 2014/08/19 & 06:57 & & 2014/08/19 & 20:30 & 0.14 & & 2014/08/19 & 23:30 & 1.5 & & DSF & 2014/08/15 & 17:09 &   -- & N26E18\\
                     74 & 2014/09/11 & 23:45 & & 2014/09/12 & 09:30 & 0.75 & & 2014/09/12 & 08:30 & 2.0 & & FLA & 2014/09/09 & 00:38 & M4.5 & N12E29\\
                     75 & 2014/09/12 & 15:53 & & 2014/09/12 & 20:30 & 1.09 & & 2014/09/13 & 01:30 & 5.1 & & FLA & 2014/09/10 & 17:33 & X1.6 & N14E02\\
                     76 & 2014/11/10 & 02:21 & & 2014/11/10 & 13:30 & 0.75 & & 2014/11/10 & 19:30 & 3.6 & & FLA & 2014/11/07 & 17:26 & X1.6 & N15E35$^{\ast}$\\
                     77 & 2014/12/21 & 19:11 & & 2014/12/22 & 12:30 & 1.82 & & 2014/12/22 & 14:30 & 6.0 & & FLA & 2014/12/17 & 04:42 & M8.7 & S20E09\\
                     78 & 2014/12/22 & 15:11 & & 2014/12/23 & 10:30 & 0.69 & & 2014/12/23 & 00:30 & 0.7 & & FLA & 2014/12/20 & 00:26 & X1.8 & S21W24\\
                     79 & 2014/12/23 & 11:15 & & 2014/12/23 & 11:30 & 1.01 & & 2014/12/23 & 20:30 & 1.7 & & FLA & 2014/12/21 & 12:17 & M1.0 & S11W28$^{\ast}$\\\hline
   \enddata
   \tablenotetext{a}{The maximum density depression in FD observed with the GMDN together with its recorded date and time.
   The maximum density depression in \% is normalized to the 6 hours average of the GCR density prior to the SSC onset.
   For our derivations of the GCR densities, see Section \ref{sec:AnaXi} in the text.}
   \tablenotetext{b}{The maximum density depression in FD observed with the NMs together with its recorded date and time.}
   \tablenotetext{c}{Type of the solar eruption specified from the SW news; FLA is flare and DSF is filament disappearance.}
   \tablenotetext{d}{Solar event properties given in the SWPC solar event list.
      Listed date and time represent the recorded time of the maximum intensity of H-$\alpha$ or X-ray emissions,
   while those indicate the start time of event for the filament disappearance.}
   \tablenotetext{\dagger}{Excluded from the analysis in this paper due to the lack of the GMDN data.}
   \tablenotetext{\ddagger}{Excluded from the analysis in this paper because the heliographic location of the solar eruption could not be specified in this event.}
   \tablenotetext{\ast}{$\rm L$ocation of the solar eruption is specified from the SWPC Solar Region Summary report,
   because the SWPC solar event list provides only the solar region number for this event.}
   \label{tb:ev}
\end{deluxetable}

\clearpage

\end{document}